\documentclass[sigconf, nonacm]{acmart}
\usepackage{graphicx}
\usepackage[utf8]{inputenc}      
\DeclareUnicodeCharacter{202F}{\,}  
\usepackage{xspace,colortbl,multirow}
\usepackage{amsmath}
\usepackage{cancel}
\usepackage{array}
\usepackage{verbatim}
\usepackage{dsfont}
\usepackage{color}
\usepackage{amssymb}
\usepackage{subfig}
\usepackage{pifont}
\usepackage{multicol}
\usepackage{hyperref}
\usepackage{multirow}
\usepackage{colortbl}
\usepackage{xcolor}
\usepackage{tcolorbox}
\usepackage{enumitem}
\usepackage{lipsum}
\usepackage{fancyvrb}
\usepackage{tcolorbox}
\tcbuselibrary{breakable,listings,skins}
\usepackage{booktabs}
\usepackage{longtable}
\usepackage{ragged2e}

\makeatletter
\newif\if@restonecol
\makeatother

\usepackage[lined,boxed,vlined,ruled,linesnumbered]{algorithm2e}
\usepackage{hyperref}
\usepackage{listings}
\usepackage{setspace}
\usepackage{tabularx}
\usepackage{array}
\usepackage{booktabs}
\usepackage{listings}
\usepackage{enumitem}
\usepackage{xltabular}
\usepackage{cuted}
\usepackage{supertabular, array}
\usepackage{multicol}
\usepackage{supertabular}
\usepackage{geometry}

\lstdefinestyle{prettysql}{
    language=SQL,
    basicstyle=\ttfamily\small,
    keywordstyle=\color{blue}\bfseries,      
    stringstyle=\color{red},                  
    commentstyle=\color{gray}\itshape,        
    identifierstyle=\color{black},            
    morekeywords={SELECT, FROM, JOIN, ON, WHERE, Aggregate, Filter, Join, Extract},
    sensitive=true,
    frame=single,                             
    backgroundcolor=\color{gray!10},          
    rulecolor=\color{blue!70},               
    numbers=left,                             
    numberstyle=\tiny\color{gray},           
    numbersep=5pt,                          
    breaklines=true,                         
    breakatwhitespace=true,                  
    tabsize=1,                               
    showspaces=false,                        
    showstringspaces=false,                  
    xleftmargin=20pt,                        
    framexleftmargin=15pt,                   
}

\newtheorem{example}{Example}

\newcommand{\sys}{\ensuremath{{\tt UDA}$-${\tt Bench}}\xspace}

\newcommand{\evaporate}{\ensuremath{\texttt{Evaporate}}\xspace}

\NewDocumentCommand{\cc}{ mO{} }{\textcolor{red}{\textsuperscript{\textit{CC}}\textsf{\textbf{\small[#1]}}}}
\NewDocumentCommand{\SZZ}{ mO{} }{\textcolor{blue}{\textsuperscript{\textit{SZZ}}\textsf{\textbf{\small[#1]}}}}
\NewDocumentCommand{\DQY}{ mO{} }{\textcolor{orange}{\textsuperscript{\textit{DQY}}\textsf{\textbf{\small[#1]}}}}
\NewDocumentCommand{\LJQ}{ mO{} }{\textcolor{green}{\textsuperscript{\textit{LJQ}}\textsf{\textbf{\small[#1]}}}}

\newcommand{\e}{\ensuremath{\mathcal{E}}\xspace}

\newcommand{\extract}{\texttt{Extract}\xspace}
\newcommand{\filter}{\texttt{Filter}\xspace}
\newcommand{\join}{\texttt{Join}\xspace}
\newcommand{\agg}{\texttt{Agg}\xspace}

\newcommand{\quest}{\texttt{QUEST}\xspace}
\newcommand{\zendb}{\texttt{ZenDB}\xspace}
\newcommand{\uqe}{\texttt{UQE}\xspace}
\newcommand{\eva}{\texttt{Evaporate}\xspace}
\newcommand{\pz}{\texttt{Palimpzest}\xspace}
\newcommand{\lotus}{\texttt{LOTUS}\xspace}
\newcommand{\docetl}{\texttt{DocETL}\xspace}

\newcommand{\nba}{\texttt{NBA}\xspace}
\newcommand{\art}{\texttt{WikiArt}\xspace}
\newcommand{\legal}{\texttt{Legal}\xspace}

\newcommand{\med}{\texttt{Healthcare}\xspace}
\newcommand{\finan}{\texttt{Finance}\xspace}

\newcommand\vldbpagestyle{plain} 

\begin{document}

\acmConference[SIGMOD' 25]{Make sure to enter the correct
	conference title from your rights confirmation email}{June 22--27,
	2025}{Berlin, Germany}

\title{Unstructured Data Analysis using LLMs: A Comprehensive Benchmark [Experiments \& Analysis]}

\setcopyright{acmlicensed}
\copyrightyear{2025}
\acmYear{2025}
\acmDOI{XX.XXXX/XXXXXXX.XXXXXXX}
\acmPrice{}

\author[]{Qiyan Deng, Jianhui Li, Chengliang Chai, Jinqi Liu, Junzhi She, Kaisen Jin, Zhaoze Sun, Yuhao Deng, Jia Yuan$^{\dagger}$,  Ye Yuan, Guoren Wang, Lei Cao$^{ *\dagger}$}
\affiliation{
    \institution{University of Arizona$^{\dagger}$, MIT$^{*}$, Beijing Institute of Technology}
}

\acmISBN{979-8-4007-0422-2/24/06}

\begin{abstract}
Nowadays, the explosion of unstructured data presents immense analytical value. 
Leveraging the remarkable capability of large language models (LLMs) in extracting attributes of structured tables from unstructured data, researchers are developing LLM-powered data systems for users to analyze unstructured documents as working with a database. These unstructured data analysis (UDA) systems differ significantly in all aspects, including query interfaces, query optimization strategies, and operator implementations, making it unclear which performs best in which scenario. Unfortunately, there does not exist a comprehensive benchmark that offers high-quality, large-volume, and diverse datasets as well as rich query workload to thoroughly evaluate such systems. 
To fill this gap, we present \sys, the first benchmark for unstructured data analysis that meets all the above requirements. Specifically, we organize a team with 30 graduate students that spends over in total 10,000 hours on curating 5  datasets from various domains and constructing a {\it relational database view} from these datasets by manual annotation. These relational databases can be used as ground truth to evaluate any of these UDA systems despite their differences in programming interfaces. Moreover, we design diverse queries to analyze the attributes defined in the database schema, covering different types of analytical operators with varying selectivities and complexities. We conduct in-depth analysis of the key building blocks of existing UDA systems: query interface, query optimization, operator design, and data processing. We run exhaustive experiments over the benchmark to fully evaluate these systems and different techniques w.r.t. the above building blocks. The major outcomes of this project, including (1) a comprehensive benchmark that allows a rigorous evaluation of UDA systems and (2) a deep understanding of the strengths and limitations of existing systems, pave the way for future research of unstructured data analysis.

\end{abstract}

\settopmatter{printacmref=false}
\renewcommand\footnotetextcopyrightpermission[1]{}

\maketitle

\pagestyle{\vldbpagestyle}

\section{introduction}\label{sec: introduction}
Modern organizations store a large quantity of unstructured data such as clinical notes, legal contracts, financial reports, etc., which account for 80\%-90\% of global data based on IDC research~\cite{idc}. These vast repositories of unstructured data, if analyzed appropriately, have immense value in various domains. 

As an example, healthcare providers often manage a corpus of hundreds of thousands of heterogeneous medical documents, including disease documents (detailing etiology, symptomatology, and progression patterns), drug documents (detailing indications, mechanisms of action, and contraindications) and documents of medical institutions. 
If there were a unstructured data analysis system, it could enable a provider to easily assist a patient who, for example, has symptoms of frothy urine and dull flank pain, by running queries to identify possible diseases, 
recommend public hospitals within certain distances of the patient’s home that specialize in treating these diseases.

\noindent \textbf{LLM-powered Unstructured Data Analysis.} To support such needs, the database community is actively developing LLMs-based systems for UDA~\cite{uqe, zendb, lotus, pz, docetl}. These systems harness LLMs to generate information from multi-modal data lakes (encompassing text, images, etc.) and perform analytical operations such as filtering, aggregation, and join, thanks to the ever growing semantic comprehension and reasoning capabilities of LLMs. 

Although these systems use different query interfaces, e.g., Python or SQL queries, they are all declarative systems which, similar to relational databases, offer a set of logical operators for users to write a simple program to analyze data. These systems provide optimized implementation of these operators to ensure accuracy and reduce LLM cost. Typically, these systems also feature an optimizer that automatically transforms the user program to an optimized query execution plan, with optimizations such as  ordering the filters, converting joins to filters, selecting an appropriate LLM for the task, etc.  In this way, these systems abstract away time-consuming engineering details in UDA, while transparently optimizing accuracy and addressing the performance and scaling bottleneck of LLMs.

\noindent\textbf{The Need of a Comprehensive Benchmark.} 
Clearly, different systems have different implementations of operators and query optimization strategies. It is unclear which system works best in which scenario. Furthermore, these systems all use small datasets and query workloads in their evaluation, making the results less convincing.  For example, ZenDB~\cite{zendb} uses merely three datasets with 221 unstructured documents and 27 queries in total, and the ground truth is not publicly available.
Palimpzest~\cite{pz} provides a text document dataset with 1,000 short emails, and only one query is available on this dataset.
DocETL~\cite{docetl} evaluates extraction tasks on five datasets with hundreds of documents, but each of them defines only one or two extractable attributes, which limits the diversity and complexity of data analysis.

Therefore, a comprehensive benchmark is in desperate need
to standardize the evaluation of LLM-powered UDA systems and guide the design of such systems.

\noindent\textbf{Design Goals.} To fill this gap, we aim to construct a comprehensive benchmark guided by the following design goals in this work:


\noindent\textit{\underline{(1) Dataset Volume.}} To thoroughly evaluate the efficiency and scalability of different methods, the benchmark has to involve unstructured datasets of various volumes, especially large-scale ones, which are measured from two aspects: the number of unstructured documents and the length of each document. The rationale is that, when dealing with large-scale datasets, there are significant challenges related to both the latency and cost of LLMs.

\noindent\textit{\underline{(2) Dataset \& Query Workload Variety.}} 
Evaluating the performance of UDA systems must take into account the properties of datasets and the query workloads.
In general, important properties w.r.t. datasets include domains, data modality, whether the documents have a clear structure, etc. The query workloads should cover different combinations of analytical operators (e.g., filter, aggregation, join) as well as various selectivities of filters.


\noindent\textit{\underline{(3) Precise labels.}}
A high-quality benchmark must provide precise labels as ground truth.  With precise labels, the benchmark could provide an accurate evaluation with respect to the performance of different systems, regardless of their different system interfaces or execution strategies. 




\noindent\textit{\underline{(4) Easy to evaluate existing systems.} }
To comprehend the advantages/disadvantages of existing UDA systems and expose research opportunities, we have to implement the above approaches reliably, perform a thorough evaluation and conduct a detailed analysis of the results.


\noindent\textbf{Our Proposal.} We construct \sys --the first benchmark for UDA that meets all the above goals.

\noindent\underline{Key Insight: A Relational Database Review.} To develop such a benchmark, the key insight is that constructing a relational database from the corresponding multi-modal data lake is sufficient to cover the evaluation needs of all existing systems, including accuracy, latency, and cost.

\begin{example}
Still using the medical application as an example, given a category of files, e.g., disease documents, we can define a relational schema (i.e., a number of meaningful attributes such as disease name, etiology, symptoms, etc.) in advance. Then we extract the values of all attributes from the data lake as the ground truth. In this way, we obtain a relational database consisting of multiple tables, each corresponding to a specific category of files (i.e., disease, drug, medical institution, etc.). Consequently, any analytical queries concerning the aforementioned attributes can be evaluated based on corresponding data records in the constructed database view, agnostic to their query semantics and execution strategies. For example, although the semantic operators in Lotus prefer natural language input, it can still compose queries about the attributes covered by the database to evaluate their implementation. 
\end{example}

Guided by this insight, we construct the benchmark to achieve all the above goals. To achieve the dataset volume goal, we collect five sets of unstructured documents, as shown in Table~\ref{tab:dataset-stats}. In terms of the number of documents, all the datasets have at least hundreds of documents and, in particular, \med has 100,000 documents, which is \textbf{\texttt{100$\times$}} more than the existing benchmarks. In terms of lengths of the documents, two of them have an average length of more than 10,000 tokens. In particular, on 
\finan dataset, the document length can be up to 838,418 tokens ($\approx$100 pages).

For the second design goal of ensuring the diversity of the datasets and query workloads, we construct datasets from various domains, including healthcare, law, art, sports, and finance. Among them, art and finance include images and text.
%
%
We construct a total of \textbf{240} queries, \textbf{\texttt{10$\times$}} more than the current benchmarks. These queries are grouped into \textbf{5} categories: \texttt{Select}, \texttt{Select+Filter}, \texttt{Select+Aggregation}, \texttt{Select+Join}, and other complex queries that are combinations of at least 3 operator types.

For the third goal of precise labels, we define \textbf{147} meaningful attributes that can be extracted from the five datasets, including categorical, numerical, and string types, exceeding the existing benchmarks by \textbf{\texttt{10$\times$}}. A team of 30 graduate students spends more than 10,000 hours on labeling, with the assistance of LLMs in cross-validation for quality assurance. 
%

To achieve the last goal of easy evaluation,
we have conducted extensive experiments to evaluate these systems on all datasets, measuring accuracy, cost, and latency across fine-grained query categories to capture differences in system behavior.
Furthermore, we analyze the experimental results from four perspectives: query interface, query optimization, operator design, and data processing, which are the key building blocks of such systems.
This in depth, multi-faceted analysis allows us to offer actionable insights.














\noindent\textbf{Contributions.} We make the following contributions: 

\noindent(1)
We present \sys, a comprehensive benchmark for unstructured data analysis that includes large-scale and diverse datasets, as well as a rich set of varied queries; to the best of our knowledge, it is the first work that constructs a comprehensive benchmark and thoroughly evaluates existing LLM-powered UDA systems. 


\noindent(2) 
We collect five sets of unstructured data from diverse domains, including more than 100,000 documents. 
We construct 240 queries involving various analytical operators over the attributes defined in the database.

\noindent(3)
We provide comprehensive, cross-validated relational tables as ground truth, annotated by 30 graduate students over 10,000 hours, enabling objective and reproducible evaluation on UDA systems.

\noindent(4)
We thoroughly evaluate seven representative UDA systems in accuracy, cost, and latency. Our in depth analysis offers insights to guide future research in this field.

\section{Unstructured Data Analysis Systems}\label{sec: system}

\begin{figure}[ht!]
    \centering
    \includegraphics[width=\linewidth]{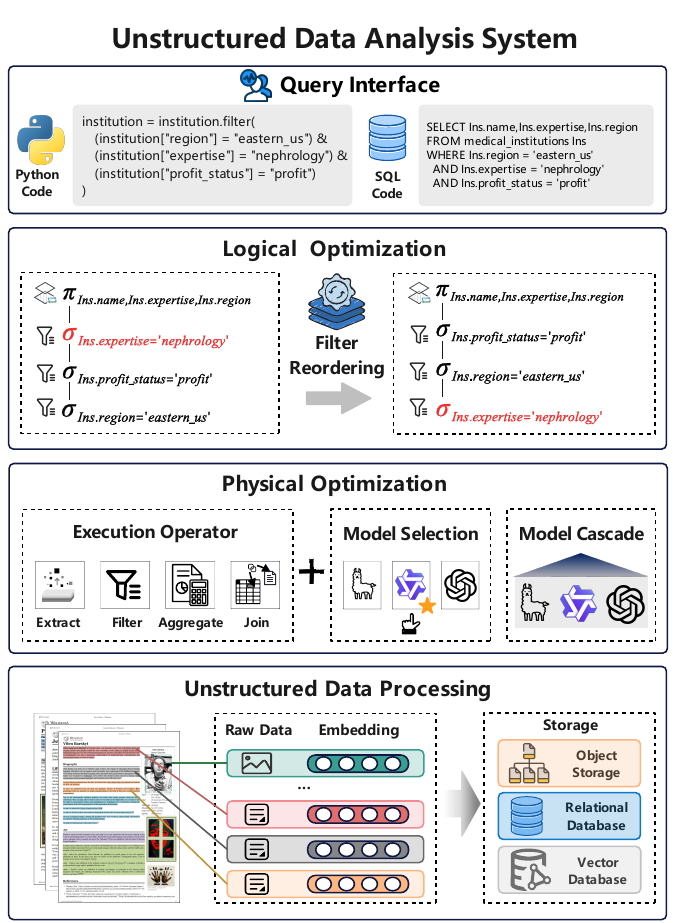}
    \vspace{-2em}
    \caption{Architecture of Unstructured Data Analysis System.}
    \vspace{-2em}
    \label{fig:UDA}
\end{figure}

In this section, we present the architecture of a typical LLM-powered data analysis system, which in general consists of 4 key modules, i.e., {\it the query interface}, {\it logical optimization layer}, {\it physical optimization layer}, and {\it data processing layer}, respectively. As shown in Figure~\ref{fig:UDA}, the user submits a query via the interface, which describes the analytical task. The system parses the query into some analytical operators, such as \extract, \filter, \join, and \texttt{Aggregation}. Then, the logical optimization layer determines an optimal execution plan, such as pushing down predicates or ordering the operators. Subsequently, there are typically multiple methods (e.g., using different types of LLMs) to implement each operator, resulting in various physical plans. The physical optimization layer alternatively selects the appropriate implements and produces an optimized physical execution plan.
Finally, the data processing layer serves as the  data foundation for the aforementioned optimizations. 
Beyond the raw data stored in distributed storage systems (e.g., Amazon S3), the systems often segment documents into chunks. These document chunks, along with images within documents, are subsequently transformed into embeddings, which are then loaded into a vector database. This infrastructure enables accurate and efficient retrieval to identify the information relevant to a query, ultimately speeding up the query execution and reducing the cost.


\noindent \textbf{System Optimization Goals.} To summarize, unlike traditional databases that focus mainly on optimizing latency, LLM-powered UDA systems have multiple optimization goals, i.e., accuracy, cost, and latency. 
Firstly, accuracy is critical in such systems for two primary reasons. (1) LLMs are prone to hallucinations, leading to potential inference errors; and (2) to reduce costs, it is a common practice to only feed the LLM  document chunks highly relevant to the analysis, rather than entire documents. However, if the retrieval misses relevant chunks or erroneously returns irrelevant chunks, it will degrade the analysis accuracy.
Secondly, LLM inference consumes substantial computational costs, which is typically calculated based on the number of input and output tokens. In UDA tasks, outputs are generally succinct, making their token cost negligible.
Consequently, reducing LLM cost is typically achieved by minimizing the number of input tokens. However, the system still has to ensure that the LLM gets sufficient information to ensure the quality of the analysis. 
Finally, LLMs face high inference latency due to numerous parameters, making it crucial to minimize query latency in data analysis systems. Inference latency largely depends on the number of input tokens, so reducing these tokens can decrease query latency, with further improvements possible through strategies like parallel processing.

We detail how existing systems use four modules to achieve the stated goals. Table~\ref{table:system_modules} shows each system's module composition.

\begin{table*}[ht]
  \centering
  \renewcommand{\arraystretch}{1.45}
  \setlength{\tabcolsep}{5pt}
  \arrayrulecolor{black}

  \resizebox{0.8\textwidth}{!}{%
    \begin{tabular}{|c|c|ccc|cccc|cc|}
      \hline
      \multirow{2}{*}{\textbf{System}} 
      & \multirow{2}{*}{\textbf{Query Interface}} 
      & \multicolumn{3}{c|}{\textbf{Data Processing}}
      & \multicolumn{4}{c|}{\textbf{Operator}}
      & \multicolumn{2}{c|}{\textbf{Query Optimization}} \\
      \cline{3-11}
      & 
      & \textbf{Chunking}
      & \textbf{Embedding}
      & \textbf{Multi-modal}
      & \textbf{\extract}
      & \textbf{\filter}
      & \textbf{\join}
      & \textbf{\agg}
      & \textbf{Logical}
      & \textbf{Physical}\\
      \hline
      \eva & \ding{55}   & \ding{55} & \ding{55} & \ding{55} & \ding{51} & \ding{55} & \ding{55} & \ding{55} & \ding{55} & \ding{55} \\
      \hline
      \pz        & Code       & \ding{55} & \ding{55} & \ding{51}  & \ding{51} & \ding{51} & \ding{55} & \ding{55} & \ding{51} & \ding{51} \\
      \hline
      \lotus     & Code       & \ding{55} & \ding{51} & \ding{51} & \ding{51} & \ding{51} & \ding{51} & \ding{51} & \ding{55} & \ding{51} \\
      \hline
      \docetl    & Code    & \ding{51} & \ding{51}  & \ding{55} & \ding{51} & \ding{51} & \ding{51} & \ding{51} & \ding{51} & \ding{51}  \\
      \hline
      \zendb     & SQL-like      & \ding{51} & \ding{55}   & \ding{55} & \ding{51} & \ding{51} & \ding{51} & \ding{55} & \ding{51} & \ding{55}  \\
      \hline
      \quest     & SQL-like      & \ding{51} & \ding{51} & \ding{55} & \ding{51} & \ding{51} & \ding{51} & \ding{55} & \ding{51} & \ding{55}  \\
      \hline
      \uqe       & SQL-like      & \ding{55} & \ding{51} & \ding{55} & \ding{51} & \ding{51} & \ding{55} & \ding{51} & \ding{51} & \ding{55}  \\
      \hline
    \end{tabular}%
  }
  \caption{Overview of Existing Unstructured Data Analysis Systems.}
          \vspace{-3em}
  \label{table:system_modules}
\end{table*}

\subsection{Data Processing}~\label{subsec:data}
In general, these systems are built to handle a wide variety of data types, including plain text, images, etc., which are extracted from complex documents such as PDF by OCR tools. Then, they organize these contents into different formats, which are further processed by different  strategies to support downstream analytical tasks. 
%


\lotus~\cite{lotus}, \uqe~\cite{uqe} and \docetl~\cite{docetl} typically organize plain text and images into semi-structured files (CSV for \lotus and \uqe, JSON for \docetl), where each entry corresponds to a document. 
For CSV, plain text is stored in the text column and the paths of the image files are recorded in the image column. For JSON, each document is represented as an object with text and image fields.
%
Subsequently, these systems transform text into embeddings for later semantic analysis, adding the storage path of these embeddings as an additional column in the CSV file.
\uqe, which supports images, stores the embedding of images as another additional column to support the aggregation operator. 


 \zendb~\cite{zendb} and \quest~\cite{quest} target analyzing plain text in documents, which is stored in a relational database. \zendb leverages the visual features of PDF such as font size, boldness, and positioning to identify hierarchical section titles and divides the document into semantic units, forming a Semantic Hierarchical Tree (SHT) that reflects the structure of the document. Then, it summarizes the content under each title (i.e., a tree node) using the NLTK toolkit~\cite{NLTK}
 and stores the summarization in the node. In addition, \zendb calculates the embedding of each sentence for a subsequent cost-effective attribute extraction. The SHT is stored in a database table, where each row corresponds to a node, recording the document ID, node name, plain text, summarization, etc.



\quest, on the other hand, constructs two levels of indexes to support accurate and cost-effective attribute extraction.  It first generates a summary for each document using the NLTK toolkit and encodes it with the E5 model to build a document-level index. This index allows the system to quickly exclude documents that are irrelevant to the query.
Then, it applies LangChain’s \textit{SemanticChunker} to split each document into semantically coherent chunks. That is, within each chunk, every two adjacent sentences have similar semantics, i.e., high similarity between their embeddings.
%
%
\sys again uses the E5 model to embed these chunks and constructs a segment-level index, which is used to identify chunks relevant to a to-be-extracted attribute. This avoids feeding the entire document to LLMs to save cost. 
The embeddings of both document summaries and chunks are stored in a vector database, while the corresponding text chunks and their metadata (e.g. plain text, embedding indices) are stored in a relational database.

\eva~\cite{eva} only supports plain text as well, which is stored in a folder for subsequent analysis. \pz~\cite{pz} organizes the plain text, images each document into a specific directory, where distinct subfolders correspond to different types of data.

\subsection{Query Interface \& Operators}~\label{subsec:query}

\noindent \textbf{Query Interface.} 
Each system provides a query interface for users to define analytical tasks on processed data.
\uqe, \zendb, and \quest use SQL-like language, using relational syntax to support analysis over unstructured data. 
For example, in the \med dataset, to find private institutions specializing in nephrology and located in the eastern United States, a user can write a query as shown in Figure~\ref{fig:UDA}.
%
\docetl, \evaporate, \lotus and \pz offer declarative Python APIs, corresponding to the logical operators in relational databases, for users to compose a query as a Python program. For example, the user query described above could be represented using code as shown in Figure~\ref{fig:UDA}.

\noindent \textbf{Query Operators.} In \sys, we include 4 common analytical operators as below.

\noindent \underline{\texttt{Extract}} aims to extract relevant attributes from a set $D$ of unstructured data.  Formally, \extract is defined as $\text{\extract}(D, A)$, where $A$ specifies the attributes to be extracted from $D$ and the output is a relational table $T_D$.
For example, to extract the rating of the institutions and their location, this operation can be expressed as \extract(\med, [\texttt{rating}, \texttt{location}]). In an SQL-like interface, this corresponds to \texttt{SELECT rating, location FROM \texttt{Healthcare}''}, while in the Python API, it could be written as \texttt{pz.addcolumns\\(\texttt{Healthcare}, [rating, location])''}.
In addition, users can provide attribute descriptions as prompts, helping LLMs produce accurate answers.
Specifically, \eva employs LLMs to generate code to extract each attribute. Other systems feed the attribute (with optional user descriptions) and relevant document chunks (possibly the entire document) to LLMs for extraction.


\noindent \underline{\texttt{Filter.}} 
Given a condition $C$,  \filter operation selects a subset of documents that satisfy $C$ from $D$, denoted by $\text{\filter}(D,C)$. 
Specifically, $C$ defines a filter on document attributes, and evaluates whether the relevant attributes satisfy the filter for each document.
%
For example, \texttt{``lotus.sem\_filter(`the \{document\} satisfy \{profit\_status\} is profit')''} is a filter in \lotus expressed with Python APIs.
In general, existing systems adopt two strategies to implement such filters: (1)  \pz,  \quest, and \zendb first extract the \texttt{profit\\ \_status} value from each document and then evaluate whether it satisfies the filter; and (2) \lotus , \docetl and \uqe take the filter as a part of prompts and leverage LLMs to determine whether the filter condition is satisfied.

\noindent \underline{\texttt{Join}} is a cross-document operation in UDA, i.e., combining information from two sets of documents based on a specified join condition $a$.
Formally, \join is defined as $\text{\join}(D_1, D_2, a)$, where $D_1$ and $D_2$ are two subsets of documents, and $a$ is the join attribute. This indicates that if the system extracts two tables (both contain the attribute $a$) respectively, the tables can be joined on $a$.
For example, 
from the document subset of \texttt{disease}, the system could extract a table describing the disease and another table with drug attributes from the \texttt{medication} subset; and the two tables can be joined by the disease name. In this way, a user can easily identify possible diseases based on symptoms and then find medications that can treat those diseases through the join operation.
%
\zendb, \quest  implement the \join by first extracting the disease table and the medication table, respectively, from two sets of documents and joining the two tables.
On the other hand, \lotus first extracts the disease table and embeds the values of the join key attribute. It then uses the embedding of each disease to retrieve relevant medication documents. From this subset of documents, it extracts the medication table. Finally, the two tables are joined to produce the join result.

\noindent \underline{\texttt{Aggregation}} is defined as $\text{\agg}(D, a, F)$, where $a$ specifies the grouping attribute and $F$ defines the aggregation functions to apply, such as \texttt{Count}, \texttt{Sum}, \texttt{Avg}, \texttt{Min} or \texttt{Max}.
The operator supports analytical tasks like ``computing the number of institutions grouped by expertise'', i.e., \texttt{\agg(\med, `expertise', count)}, which can be represented as \texttt{``pz.GroupBySig(\med, count,`expertise')''} using  the Python API of \texttt{PALIMPZEST}.
%
To achieve this, \evaporate, \zendb, \quest and \pz extract the \texttt{`expertise'} from all documents, group the values in a table, and then perform aggregation on the grouped data. 
\lotus, \uqe, and \docetl, on the other hand, first preprocess the documents by clustering them based on their embeddings, and then perform aggregation or batch inference within each cluster as an approximation to save costs.
Besides, \uqe supports grouping images according to their embeddings.

\subsection{Logical Optimization}~\label{subsec:logical}
Given a user-specified query involving multiple operators, UDA systems generate an optimized logical plan to reduce LLMs costs and query latency. The optimizations adopted by existing systems mainly include filter reordering, filter pushdown, and join ordering.



\noindent \textbf{Filter Reordering.}
Consider a query that selects artists and their birthdates with two filters, i.e., $F_1=$\texttt{filter(D, ``lifespan is less than 35'')} and $F_2=$\texttt{filter(D, ``tone is warm''))}. Different systems employ different optimization strategies for reordering filters to reduce costs.

\noindent\textit{\underline{(1) Selectivity-only Strategy.}} 
\pz and \uqe adopt a selectivity-only filter reordering strategy that prioritizes a filter with low selectivity. This indicates that the attribute values that a document contains have a small probability to satisfy the predicates of this filter. This reduces the chance of extracting other attributes and evaluating the corresponding filters. These systems estimate the selectivity (denoted by $sel()$) by sampling. 
%
%
For example, if $sel(F_1)=0.2$ and $sel(F_2)=0.1$, applying $F_2$ first would leave $\approx$10\% documents for $F_1$ to evalute, potentially significantly reducing the cost. However, only considering the selectivities tends to be suboptimal in minimizing the cost. For example, although $sel(F_2) < sel(F_1)$, if the document chunks that $F_2$ has to examine contain much more tokens than those of $F_1$, using this order might lead to higher costs than applying $F_1$ first.


\noindent\textit{\underline{(2) Selectivity-cost strategy.}} 
To address the above limitation, \zendb ranks filters based on scores computed by $sel(F) \times cost(F)$ and prioritizes those with lower scores. To be specific, given the attribute $a$ of a filter $F$ and an unstructured document $d \in D$, we use $d[a]$ to denote the chunk(s) that are highly relevant to $a$.  The relevance can be computed by measuring the similarity between the embeddings of chunks and the attribute. We use $c(d[a])$ to denote the number of tokens of $d[a]$. Then in \zendb, $cost(F)$ corresponds to the average number of tokens of chunks relevant to $a$ across all documents, i.e., $cost(F) = \frac{\sum_{d\in D} c(d[a])}{|D|}$. 
Therefore, \zendb produces one single filter order with respect to all documents, similar to traditional databases. This is a coarse-grained optimization because different documents might contain different numbers of tokens with respect to an attribute.

Leveraging this optimization opportunity, \quest produces different orders for different documents considering both the selectivity and the cost of every document, abandoning the above ``one single order for one query'' strategy. 
Therefore, \quest does not have the $cost(F)$ for the overall document set, but instead, for each $d \in D$,it uses $cost_{d}(F) = c(d[a])$ to denote the number of tokens w.r.t. the attribute $a$ in $d$.
Then, each document $d$ should follow a specific optimal order which prioritizes the filter with lower values of $sel(F) \times cost_{d}(F)$.

\noindent \textbf{Filter Pushdown.}
Given a query that joins two sets of documents with filters applied on them, the most straightforward way (\zendb) is to first pushdown the filters to the two document sets respectively, extract the join key attribute, and then join. Traditional databases adopt this strategy because filters typically have a lower time complexity than joins and thus should be prioritized. However, in this unstructured data analysis scenario where LLM cost is the primary optimization goal, a join may not be more costly than a filter and potentially could have a higher priority. 
Inspired by this insight, \quest proposes a join transformation strategy that first extracts the join attribute of one table and then uses the extracted values as filters to filter the other table, i.e., transforming a join into a filter. Treating this automatically generated filter equally to other filters, the \quest optimizer uses the cost model discussed above to order these filters. In this way, \quest might prioritize joins over filters to minimize the LLM cost, contradictory to the filter pushdown principle in traditional databases.
%
%

\noindent \textbf{Join Ordering.} For multi-join, \zendb and \quest dynamically and progressively decide the join order during query execution. More specifically, they first selects two tables to join based on their cost model, and it will determine the next join only after the first join finishes execution. This process iterates in a left-deep manner until all joins have been executed.

\subsection{Physical Optimization}~\label{subsec:physical}
As discussed above, each logical plan consists of a sequence of operators. Subsequently, the systems have to select an appropriate implementation w.r.t. each operator based on the properties of the data and user preferences, i.e., physical optimization. Next, we introduce the typical physical optimization strategies in UDA.


\noindent \textbf{Model Selection.} For each logical operator, the optimizer selects the most suitable model from a set of candidate based on users' preference, e.g., accuracy or cost. 
%
In \pz, if a user wants to achieve target accuracy while at a relatively low cost, the system prefers a lightweight model for simple extraction tasks to reduce cost, e.g., using \texttt{GPT-4.1-mini} to extract \texttt{``birthdate''} from the \art dataset. Conversely, for complex semantic analysis where \texttt{Llama} fails to meet the target accuracy, it employs more advanced models, such as using \texttt{GPT-4.1} to infer whether a case is a first-instance trial in the \legal dataset.

\noindent \textbf{Model Cascade.} In addition to selecting one model that is the most suitable for the operator, the optimizer in \lotus applies the model cascade technique to save more cost while meeting the users' target accuracy. More specifically, it uses a sequence of different models to execute an operator. These models have various characteristics, such as diverse qualities, varying costs, and different levels of latency.
For example,  the model cascade can be \{\texttt{GPT-4.1-nano}, \texttt{GPT-4.1-mini}, \texttt{GPT-4.1}\}, which begins with the cheapest  \texttt{GPT-4.1-nano}. \lotus immediately returns the result if the Llama output meets the target accuracy. If not, the input proceeds to the next model in the cascade, i.e., \texttt{GPT-4.1-mini}, until obtaining a satisfactory output or reaching the last model.

\noindent \textbf{Operator Decomposition.}
For each operator, \docetl pre-defines several possible decomposition strategies, each breaking down the operator into finer-grained steps to improve accuracy.
For example, it uses two decomposition strategies to implement the operator $\text{\extract}(D,[a_1,a_2])$: (1) $\texttt{split}(D,k)$ $\rightarrow$ $\texttt{reduce}$ \{ $\text{\extract}(D_1,[a_1,a_2])$, $\cdots$, $\text{\extract}(D_k,[a_1,a_2])$\}, which splits $D$ into $k$ chunks and performs joint attribute extraction on each chunk.  (2) \extract$(D,a_1)$, \extract$(D,a_2)$, which extracts $a_1$ and $a_2$ separately over the entire data set $D$ without chunking.
%
Then \docetl executes each strategy on a small validation set and uses a validation agent to evaluate and compare the quality of the output. The best-performing strategy is selected to replace the original operator in the pipeline.

\noindent \textbf{Parallel Execution.}
Leveraging efficient batched inference with vLLM~\cite{vllm},  \lotus and \docetl process multiple documents concurrently, allowing efficient operator execution over large-scale document collections.

\begin{table}[!htbp]
  \centering
  \renewcommand{\arraystretch}{1.2}   
  \setlength{\tabcolsep}{7pt}         
  \resizebox{\linewidth}{!}{%
    \begin{tabular}{|c|c|c|c|c|}
      \hline

      \textbf{Dataset} & 
      \textbf{\#Attributes} & 
      \textbf{\#Files} & 
      \textbf{Tokens (Max/Min/Avg.)} & 
      \textbf{Multi-modal} \\
      \hline
      WikiArt    & 19  & 1,000     & 1,665 / 619 / 789       & \ding{51} \\
      NBA        & 28  & 225       & 51,378 / 73 / 8,047     & \ding{55} \\
      LCR        & 19  & 566       & 45,437 / 340 / 5,609    & \ding{55} \\
      Finance    & 30  & 100       & 838,418 / 7,162 / 130,633 & \ding{51} \\
      Healthcare & 51  & 100,000   & 63,234 / 2,759 / 10,649 & \ding{55} \\  
      \hline
    \end{tabular}
  }
  \caption{Statistics of datasets.}
          \vspace{-3em}
  \label{tab:dataset-stats}
\end{table}

\section{The Benchmark Construction}\label{sec: benchmark design}
In this section, we first overview the construction process of \sys and then introduce each step in detail.

\subsection{Overview}\label{subsec: benchmark construction process overview}

Our benchmark consists of 5 datasets, which are \nba, \art, \legal, \med and \finan. We first collected and pre-processed the raw data. Then we followed a semi-automatic, iterative process to define the attributes of the relational tables. 
After that, we spent a huge amount of human effort labeling the ground truth, i.e., the attribute values that could be extracted from the datasets, applying cross-validation and iterative prompt tuning methods. 
Finally, we manually designed 5 query templates w.r.t. each dataset and generated queries using Python scripts for benchmark evaluation.

We support the benchmark result that, each raw dataset, the benchmark associates it with a JSON file containing processed data, where different modalities within the same document are stored in a single field across various entries.
In this way, if a user wants to test her system using \sys, she can directly download the processed data, load the data into the system, run her queries, and compare the results with the ground truth table stored in the relational databases.



\subsection{Datasets}\label{subsec: Benchmark datasets}

\noindent \sys consists of five datasets with their statistics summarized in Table~\ref{tab:dataset-stats}. Next, we describe these datasets below. For more details, please refer to Appendix.

\noindent\underline{\nba} dataset is crawled from Wikipedia\cite{nba_wikipedia} that contains information about NBA including players, teams, team owners, etc., from the 20th century to the present, covering their basic and statistic information like player personal honors, team founding year, owner nationality etc.

\noindent\underline{\art} is collected from \art.org~\cite{artorg2025}, which covers artists and their artworks spanning from the 19th to the 21st centuries.  For each artist document, it includes biographical information, artistic movement, a list of representative works, and high-resolution images of them as metadata.

\noindent\underline{\legal} is sourced from AustLII~\cite{austlii} with 570 professional legal cases from Australia between 2006 and 2009, covering different types such as criminal and administrative. Each case document typically includes evidence, charges, legal fee, etc.

\noindent\underline{\finan} are collected from the Enterprise RAG Challenge~\cite{enterpriserag2024}, containing annual and quarterly financial reports published in 2022 by 100 listed companies worldwide with an average token length of 130,633. Each record typically includes mixed types of content like company name, net profit, total assets, etc. 

\noindent\underline{\med} is obtained from MMedC~\cite{mmbench}, with numerous healthcare documents since 2020. This dataset contains a massive amount of files (100,000), each file having 10,649 tokens on average. It covers various types of healthcare information, like drugs, diseases, medical institutions, news, interviews crawled from large-scale web corpora and open-access healthcare websites. 

\noindent \textit{\underline{Summarization.}} These datasets show various characteristics. Compared to other three datasets, the \nba and \art datasets are less complex due to their brevity and well-defined structure. 
\legal is more complicated because it is a domain-specific dataset containing multiple attributes that require semantic deduction to extract. \finan is another complex domain-specific dataset. The key challenge it introduces is the length of the documents, which can be up to 100 pages. Lastly, \med has the largest number of documents, containing rich information, such as healthcare advertisements. \med and \nba contain multiple categories of files, e.g., disease, drug, medical institution in \med, which can support join queries. \finan and \art are multi-modal datasets covering images. We define attributes over these images to verify the capability of systems in handling images. 

\noindent\textbf{Data processing.} We design a unified data preprocessing pipeline to handle datasets with various features. 
For each dataset, we first collected the raw data and utilized the MinerU~\cite{mineru} toolkit to parse the data when dealing with complex formats such as PDF (e.g., the \finan dataset). Then, we organized the dataset into a JSON file, where each object corresponds to an unstructured document with multiple fields including text, image URL, and metadata.

In particular, for the \med and \nba datasets, we divide the documents into multiple related categories but with different topics, each of which corresponds to a relational table. For \med, originally the dataset contains 680,000 documents about heathcare information, from which we sampled 100,000 documents. Then, we leveraged LLMs to read these sampled documents and identified the three major categories of files (6,100 documents in total), i.e., disease, drug and medical institution. The remaining files are mostly  categories 
like medical devices, health policy, etc.
For \nba, we identify 4 related document sets with different categories (i.e., NBA players, NBA teams, team managers, cities).

\subsection{Ground Truth Labeling}\label{subsec: ground truth labeling}

To label the ground truth, we first identify a number of significant attributes from each dataset or file category and then manually extract their values.

\noindent\textbf{Attributes Identification.} We hire 6  Ph.D. students from different majors (e.g., finance, law, medical) in our university to read these documents carefully and identify significant attributes that pose different levels of challenges to extract. For example, the attribute \texttt{Judge\_name} can be easily identified since it can usually be found at the beginning of each document in \legal. 
In \finan, the attribute values of \texttt{Business\_cost} vary among different industries such as raw materials and wages for car manufacturers, versus product sourcing and logistics for supermarkets. Hence, a labeler has to determine the relevant costs according to the context, extract their values, and aggregate them.
In addition, image files also contain easy to extract attributes like \texttt{Tone}, whose answers often fall into ``Neutral'', ``Bright'' and ``Dark'' that can be easily categorized. Difficult attributes, such as ``Style'', require art expertise to correctly extract.



\noindent\textbf{Labeling.} To ensure high-quality ground truth for \sys, we hire a total of 30 graduate students to manually label these attributes, spending approximately 10,000 human hours. 
Moreover, to ensure labeling quality, 
we also utilize multiple LLMs (e.g., \texttt{Deepseek-V3}, \texttt{GPT-4.1}, \texttt{Claude-sonnet-4}) to extract attributes and ask humans to double-check the manually labeled ground truth based on the results provided by LLMs. 
However, for the large-scale dataset \med, it is impractical to manually label all the ground truth. 
Therefore, we adopted a semi-automated iterative labeling strategy. Recap that in Section 3.2, \med  contains a large number of documents belonging to categories other than the three major ones mentioned above. Therefore, these documents rarely include the entities in the major categories. Consequently, we ask LLMs to analyze the 10,000 documents to label whether each of them belongs to the major categories (finally we obtain 6,100 documents). If so, we ask humans to label the attributes; otherwise, the ground truth is \texttt{NULL}.

\subsection{Query Construction}\label{subsec: Query Consturction}
In general, we first ask human experts to design meaningful query templates, and then automatically instantiate these templates with different predicate values and join conditions to construct diverse queries. In total, we created 240 queries, including 220 single-table queries and 20 multi-table queries based on the templates over the 5 datasets.

To be specific, we ask the PH.D. students to design 5 query templates per dataset based on real-world scenarios. These templates are in the form of SQL-like queries and Python code, which can be utilized to test systems with different interfaces, like \zendb and \pz. We list all the query templates in Appendix. An example is shown as below.

\begin{table}[htbp]
\centering
\begin{minipage}{0.96\linewidth}
\begin{lstlisting}[style=prettysql]
SELECT {Attribute}(s), {agg_func}({Attribute}(s))
FROM diseases
JOIN drug ON disease.name = drug.disease
WHERE disease.symptoms{operator}{literal} 
GROUP BY {group_by};
\end{lstlisting}
\end{minipage}
\caption*{A SQL Template Example.}
        \vspace{-3em}
\end{table}

\begin{table}[htbp]
\centering
\begin{minipage}{0.96\linewidth}
\begin{lstlisting}[style=prettysql]
disease_doc = disease
drug_doc = drug
disease_doc = Filter(disease_doc, disease.symptoms{operator}{literal})
disease_drug = Join(disease_doc, drug_doc, disease_doc.name = drug_doc.disease)
result = Extract(disease_drug, {Attribute}(s))
result = Aggregate(result, {group_by}, {agg_func}({Attribute}(s))
\end{lstlisting}
\end{minipage}
\caption*{A Code Template Example.}
\vspace{-3em}
\end{table}

The expert first identifies a real-life scenario (e.g., a user wants to identify disease based on his symptoms and find appropriate drugs for treatment), and then uses SQL-like queries and Python code to build a template and leaves some placeholders (i.e., \texttt{\{literal\}}, \texttt{\{group\_by\_attribute\}}). Next, we populate the placeholders according to their roles in the template to generate various queries. 
Taking the SQL-like query as an example:
(1) For the \texttt{SELECT} clause, we randomly sample from all available attributes for the population. 
(2) The \texttt{FROM} clause does not need a population, as all relevant tables are specified in the template.
For \texttt{WHERE} clause, we randomly select the filter attribute and operator (i.e., $\leq, =, \geq$) with equal probability when generating each query.
Literal values are sampled to vary the selectivities. 
(4) For \texttt{AGGREGATION} clause, we use categorical attributes for grouping and numerical attributes for aggregation, and each aggregation operator (e.g., \texttt{AVG}) is randomly chosen with equal probability when constructing queries.
(5) For \texttt{JOIN} clause, we explicitly define the join graph to guide the construction of queries. For example, in \med, valid join paths include Disease $\bowtie$ Drug, along with their corresponding join keys (e.g., Disease.name = Drug.disease). Similarly, in \nba, we construct join paths such as Players $\bowtie$ Teams $\bowtie$ Managers, together with the specific join keys that link these tables.

Moreover, we varied the number of filters in \texttt{WHERE} clause to enhance the diversity of the templates. To be specific, we control the frequencies of queries with different numbers of filters ranging from 1 to 5, corresponding to 20\%, 30\%, 30\%, 10\%, and 10\% percent of all queries, respectively. This allows us to thoroughly evaluate the systems with query in different levels of complexity. In addition, we define eight query types, ranging from simple \extract queries to more complex forms such as $\text{\extract}+\text{\filter}+\text{\agg}$, covering a broad range of real-world analytical scenarios.

\section{evaluation}\label{sec: evaluation}
In this section, we evaluate existing systems on \sys and analyze the results, aiming to answer the following questions. 

\noindent \textbf{- RQ 1}: What is the accuracy of different systems when evaluated on the benchmark?

\noindent \textbf{- RQ 2}: What is the cost of different systems on the benchmark?

\noindent \textbf{- RQ 3}: How efficient are different systems on the benchmark?

\noindent \textbf{- RQ 4}: How do different logical optimization strategies perform in such systems? 

\noindent \textbf{- RQ 5}: How do different physical optimization strategies perform?  

\subsection{Experimental Settings}

\noindent \textbf{Systems for Evaluation}. Our benchmark evaluates 7 existing unstructured data analysis systems as below. 
(1) 
\eva is a table extraction system. In our evaluation, we employ \eva to extract structured tables from documents, and subsequently execute SQL queries on the resulting tables.
(2) 
\pz provides Python API-based operators for unstructured data processing. We convert each SQL query into the corresponding \pz code, execute it and obtain the results.
(3) 
\lotus also provides an open-source Python library, which we use to execute queries through its interface.
(4) 
\docetl is an open-source project allowing users to execute queries by writing Python code. We rewrite our queries with \docetl library and execute the Python programs.
%
(5) 
\quest provides SQL-like query interface for processing unstructured documents, and we directly use their code to execute queries.
(6) 
\zendb does not provide their code; therefore, we implement their SHT chunking and filter reordering strategies and evaluate them on \sys.
(7) 
\uqe also does not provide their code; therefore, we implement its \filter\ and \agg\ operators, as well as its logical optimizations, and then execute them on \sys.
For a comprehensive evaluation, we adapted and modified these systems to support our evaluation (details are provided in Appendix).  We also list all the evaluation prompts in Appendix.

\noindent \textbf{Evaluation Metrics.}
Following existing works~\cite{quest,zendb}, we measure accuracy, cost, and latency with respect to all queries. 
For accuracy, we follow \cite{quest} and report the average precision, recall, and F1-score across all queries.
Given a query $Q$, the set of tuples returned by a method is denoted as $T(Q)$, and the ground truth is denoted as $GT(Q)$.  For each element $t\in T(Q)$, we evaluate whether it can be matched to a corresponding tuple in $GT(Q)$.  
Therefore, we have $P = \frac {|T(Q)\cap GT(Q)|}{|T(Q)|}$, $R = \frac {|T(Q)\cap GT(Q)|}{|GT(Q)|}$, $F1 = \frac {2\times P\times R}{P+R}$.
%
For LLM costs, we report the average number of tokens (in thousands) per document per query.
For latency, we report the mean execution time in seconds per document per query.

\noindent \textbf{Environment}. We implemented all experiments in Python~3.7 and run experiments on an Ubuntu Server with four Intel(R) Xeon(R) Gold 6148 2.40GHz CPUs with 80 cores in total, two NVIDIA GeForce 4090 GPUs, 1TB DDR4 main memory, and 6TB SSD. The same environments ensure a fair comparison over different methods.
We adopt \texttt{GPT-4.1-mini} as the default LLMs for API calls, and \texttt{Qwen3-Embedding-0.6B} as the default embedding model. 

\begin{figure*}[ht!]
    \centering
    \includegraphics[width=0.9\linewidth]{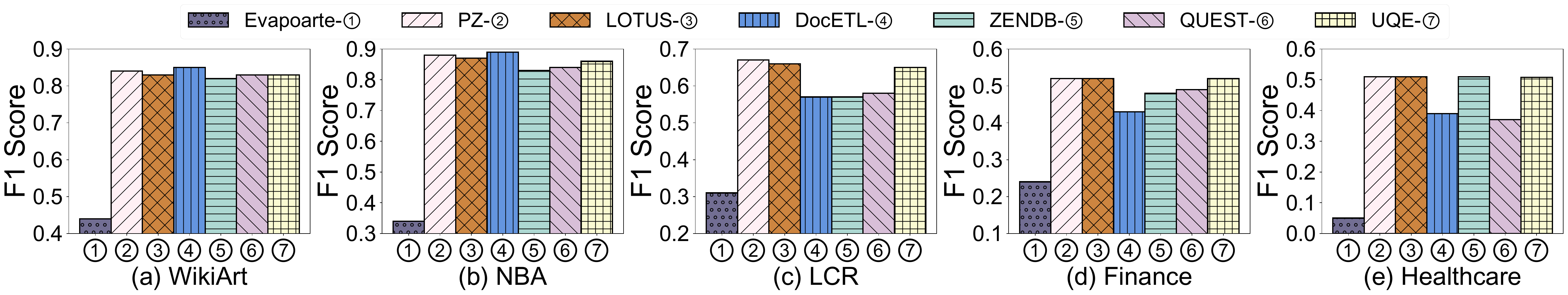}
    \vspace{-1.5em}
    \caption{F1-score Comparison of Queries with only Extraction.}
    \vspace{-1.5em}
    \label{fig:sf-f1}
\end{figure*}

\begin{figure*}[ht!]
    \centering
    \includegraphics[width=0.9\linewidth]{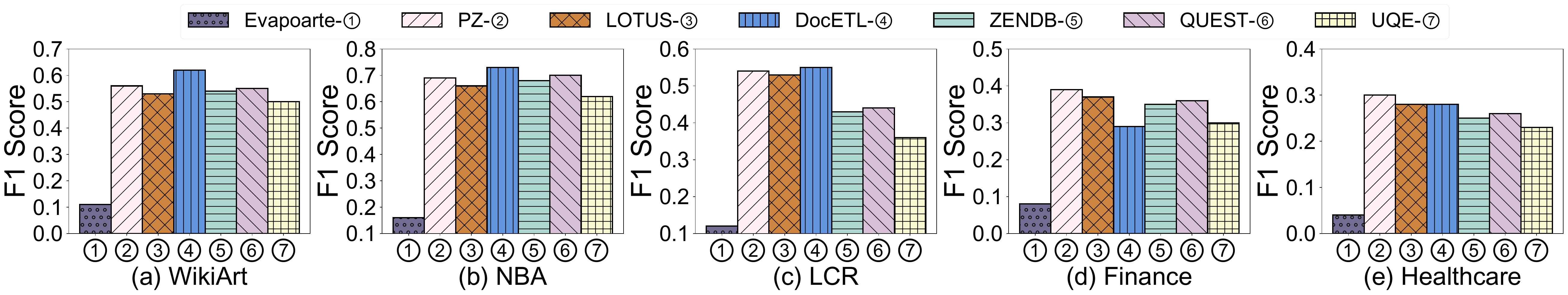}
    \vspace{-1.5em}
    \caption{F1-score Comparison of Queries with Filter.}
    \vspace{-1.5em}
    \label{fig:sfw-f1}
\end{figure*}

\subsection{Overall Accuracy Comparison (RQ1)}
\label{subsec:effectiveness}
Figure~\ref{fig:sf-f1} shows the effectiveness of extraction only queries. Almost all systems perform well on datasets \art and \nba, with an F1-score around 0.85. This is because the two datasets are relatively short or contain easy-to-extract attributes. In particular, \docetl performs the best because it leverages multi-agent techniques to extract the attributes.
\eva performs the worst because it uses LLM-generated code to extract data. However, the code essentially corresponds to a limited number of rules, which tend to be less effective when handling complex documents.

On datasets \legal, \finan, and \med, which include long documents and  multiple challenging attributes, the accuracy varies across different systems. Specifically, \lotus, \uqe, and \pz achieve similar performance and perform the best, with an accuracy around 0.66 on \legal, 0.52 on \finan and 0.51 on \med, as they all feed the entire document to LLMs and fully leverage the models’ in-context reasoning abilities. This achieves more accurate answers. 
%
Besides,  \docetl always selects the plan that splits documents into chunks, feeds each one to the LLM for extraction, produces multiple candidate answers per attribute and finally aggregates them. It is less effective on \med and \finan because long documents introduce many noisy chunks, leading to incorrect extractions.
%
For most datasets, \quest and \zendb perform worse because they only provide relevant chunks to LLMs to save cost. However, chunking often misses relevant information, leading to lower accuracy. The exception is on \med, which lacks structured information. Therefore, \zendb feeds the entire document to LLMs, resulting in performance similar to that of \lotus, \uqe, and \pz.
All datasets do not perform well on \med because many irrelevant information are extracted from the documents of other categories.
%
%

\begin{center}
\vspace{-1em}
\fcolorbox{black}{gray!10}{\parbox{\linewidth}{Summary I: For simple datasets with short content and easy-to-extract attributes, almost all systems perform well. For complex datasets, the systems that feed entire documents to LLMs perform the best because chunk-based strategies may miss relevant information. }}
\end{center}

Figure~\ref{fig:sfw-f1} shows the effectiveness of filter queries. 
Existing systems implement filters in two ways. Given a filter, 
\lotus and \docetl feed the description of the filter together with the document into an LLM and ask it to directly output a boolean output.
Other systems first extract the attribute w.r.t. the filter and then determine whether the extracted value satisfies the filter. 
We observe that \lotus performs worse than \pz, \zendb, \quest on  \art and \nba, as extracting the relevant information before evaluating the filter leads to more accurate answers. \docetl still achieves the best performance due to its multi-agent strategies. On complex datasets, \lotus outperforms many systems because it feeds the entire document into LLMs, whereas the performance of \docetl declines due to the imperfect chunking strategy.
\uqe performs worse than other systems because it samples a subset of documents to train a regression model, which is then used to predict whether the remaining documents satisfy the filter.

\begin{center}
\vspace{-1em}
\fcolorbox{black}{gray!10}{\parbox{\linewidth}{Summary II: For the filter operation, extracting the attribute first and evaluating the filter thereafter lead to more accurate results than directly executing it with LLMs. This is because decomposing the filter into the above two steps provides LLMs with more explicit instructions.}}
\end{center}

\subsection{Overall Cost Comparison (RQ2)}
\label{subsec:cost-analysis}
Table~\ref{table: extract} shows the cost of extraction only queries. On the dataset \art with short documents, systems that use chunking strategies (i.e., \quest, \zendb, and \docetl) consume more tokens than simply processing entire documents (\uqe). 
The reason is that for each attribute, \quest and \zendb select attribute-related chunks. Given multiple attributes, there tends to be a number of duplicated chunks especially when the documents are short. As a result, the total token consumption exceeds that of simply processing the full documents.
%
%
\docetl incurs the highest cost  because (1) it examines all the chunks. When precessing each chunk, it also feeds adjacent chunks into LLMs; and (2) require executing all possible plans on sampled documents to select the optimal one. 
\pz and \lotus also process entire documents, but the chain-of-thought mechanism in \pz leads to more output tokens, while \lotus processes each attribute separately, multiplying the cost by the number of attributes.
\begin{center}
\vspace{-1em}
\fcolorbox{black}{gray!10}{\parbox{\linewidth}{Summary III: For datasets with short documents, strategies that feed the entire document to LLMs and extract attributes all at once are the most cost-effective.}}
\end{center}

On datasets \nba, \legal, and \finan with long documents, we can observe  significant differences in cost across different systems.
\docetl incurs the highest cost because of feeding a number of repeated chunks into the LLM and evaluating  possible execution plans. \lotus follows, as it repeatedly feeds the entire document into the LLM for multiple attributes.
\pz and \uqe are next, as both feed the entire document to the LLM. \quest and \zendb, which employ chunking strategies, cost less. In particular, \quest is more cost-effective than \zendb, as it feeds more fine-grained chunks into the LLM, further reducing the cost.
\eva is the most cost-efficient strategy because it only uses the LLMs to analyze a small number of documents for code generation and then runs the code for extraction without additional LLM calls.
\begin{center}
\vspace{-1em}
\fcolorbox{black}{gray!10}{\parbox{\linewidth}{Summary IV: For datasets with long documents, strategies that retrieving attribute-related chunks instead of scanning entire documents are more cost-effective without sacrificing accuracy much.}}
\end{center}

Table~\ref{table: filter} shows the cost of filter queries. \uqe incurs lower cost than other systems on most datasets because it trains a regression model to determine whether each filter condition is satisfied. 
Chunking-based systems like \zendb and \quest reduce token usage compared to \pz, \lotus, and \docetl by feeding only relevant chunks to LLMs and using logical optimizations, such as prioritizing low selectivity and computational cost filters. Datasets \art, \legal, and \med lack clear structures, leading to larger chunks in \zendb and potential duplication when processing multiple attributes.
\pz is more cost-effective than \lotus and \docetl because it employs a logical plan that prioritizes filters with low selectivity.
\begin{center}
\vspace{-1em}
\fcolorbox{black}{gray!10}{\parbox{\linewidth}{Summary V: For queries with filters, \quest and \zendb apply logical optimization and chunking-based strategies to reduce costs while maintaining accuracy.}}
\vspace{-.5em}
\end{center}

Overall, based on the above two sets of experiments, we observe that on complex datasets, all systems have a relatively low accuracy (e.g., 0.5-0.6 F1-score) and incur high cost. The reasons are two-fold. 
(1) The documents are long, including much noisy information that misleads LLMs, while the chunk strategies are not perfect.
(2) Some of the attributes are difficult to extract (e.g., first\_judge, which indicates whether a judgement was the first judgement), as it might involve analyzing multiple chunks and employing inferential reasoning. This reveals research opportunity as below.
\begin{center}
\vspace{-1em}
\fcolorbox{black}{gray!10}{\parbox{\linewidth}{Opportunity I: 
A promising direction is to investigate high-quality and cost-effective strategies for extracting difficult attributes in long documents. One key problem is to design a sophisticated chunking approach to produce chunks that contain precise information w.r.t. the to-be-extracted attribute.
}}
\vspace{-.5em}
\end{center}

\subsection{Overall Latency Comparison (RQ3)}
\label{subsec:latency-analysis}
Table~\ref{table: extract} and~\ref{table: filter} compare the query latency of different systems.
We observe that \docetl is the most time-consuming, as it applies the multi-agent technique that calls LLMs multiple time, resulting in the highest overall token usage.
\lotus is the next because it calls LLMs multiple times and each time it feeds the entire document into LLMs. 
\pz follows because it incorporates the chain-of-thought mechanism, leading to more outputs.
\uqe is relatively efficient but still processes entire documents for each call. 
However, when filters are applied, its regression model significantly improves the efficiency.
\quest and \zendb are efficient because they consume fewer input tokens and trigger fewer LLM calls. \eva is the most efficient because it generates code for extraction.

\begin{center}
\vspace{-1em}
\fcolorbox{black}{gray!10}{\parbox{\linewidth}{Summary VI: Overall, latency is closely related to the input \& output tokens as well as the number of LLM calls. Hence, chunking-based strategies with logical optimization (i.e., \quest and \zendb) are the most efficient because they reduce both the number of input tokens and LLMs calls. }}
\end{center}


\arrayrulecolor{black}
\begin{table*}[!htbp]
\centering
\renewcommand{\arraystretch}{1.2}
\setlength{\tabcolsep}{5pt}
\resizebox{0.8\textwidth}{!}{%
\begin{tabular}{|c|cc|cc|cc|cc|cc|}
\hline
\textbf{Method} 
& \multicolumn{2}{c|}{\textbf{WikiArt}}
& \multicolumn{2}{c|}{\textbf{NBA}}
& \multicolumn{2}{c|}{\textbf{LCR}}
& \multicolumn{2}{c|}{\textbf{Finance}}
& \multicolumn{2}{c|}{\textbf{HealthCare}} \\
\cline{2-11}
& Cost & Latency 
& Cost & Latency 
& Cost & Latency 
& Cost & Latency 
& Cost & Latency \\
\hline
\eva    & - & 0.16 & - & 0.30 & - & 0.25 & - & 3.50 & - & 0.52 \\
\pz     & 1.64 & 1.34 & 6.67 & 1.43 & 6.96  & 1.39  & 138.90 & 8.38   & 10.80 & 2.36  \\
\lotus  & 2.33 & 1.47 & 12.41 & 1.75 & 9.93  & 1.43  & 297.85  & 13.24   & 25.10 & 3.28  \\
\docetl & 7.04 & 15.86 & 55.53 & 79.05 & 154.26 & 270.88 & 818.10 & 1509.08 & 184.30 & 304.96 \\
\zendb   & 1.94    & 1.18    & 3.72    & 0.98     & 3.92    & 0.94   & 32.33  & 3.55       & 10.31    &  1.92   \\
\quest   & 1.20 & 0.89 & 2.06 & 0.92 & 3.09 & 0.89  & 29.30 & 2.65 & 4.70 & 1.72 \\
\uqe     & 0.92 & 0.83 & 5.96 & 1.03 & 6.12 & 1.04  & 124.02 & 5.83 & 10.09 & 1.91 \\
\hline
\end{tabular}
}
\caption{Cost and Latency Comparison for Extraction Queries.}
\vspace{-2.5em}
\label{table: extract}
\end{table*}

\arrayrulecolor{black}
\begin{table*}[!htbp]
\centering
\renewcommand{\arraystretch}{1.2}
\setlength{\tabcolsep}{5pt}
\resizebox{0.8\textwidth}{!}{%
\begin{tabular}{|c|cc|cc|cc|cc|cc|}
\hline
\textbf{Method} 
& \multicolumn{2}{c|}{\textbf{WikiArt}}
& \multicolumn{2}{c|}{\textbf{NBA}}
& \multicolumn{2}{c|}{\textbf{LCR}}
& \multicolumn{2}{c|}{\textbf{Finance}}
& \multicolumn{2}{c|}{\textbf{HealthCare}} \\
\cline{2-11}
& Cost & Latency 
& Cost & Latency 
& Cost & Latency 
& Cost & Latency 
& Cost & Latency \\
\hline
\eva    & - & 0.16 & - & 0.30 & - & 0.25 & - & 1.50 & - & 0.52 \\
\pz     & 2.35 & 1.46   & 11.77 & 6.23  & 10.70 & 4.75   & 213.80 & 12.71  & 18.30 & 7.16  \\
\lotus  & 4.70 & 6.78   & 11.91 & 6.61  & 14.89 & 4.25   & 216.10 & 18.49  & 18.40 & 7.81  \\
\docetl & 6.15 & 13.71  & 52.92 & 74.86 & 32.70 & 31.75 & 184.06 & 14.67 & 113.35 & 47.06 \\
\zendb   & 3.51    & 2.13      & 9.59  & 4.56     & 26.69  & 27.72   & 49.18 & 4.73         & 19.50     & 2.28     \\
\quest   & 2.90 & 1.86   & 5.60  & 2.12  & 9.33  & 2.10   & 36.00  & 4.33   & 9.46  & 2.10  \\
\uqe     & 0.97 & 0.74   & 4.81 & 0.99  & 8.00 & 1.32   & 33.27 & 3.98  & 8.50 & 1.34  \\
\hline
\end{tabular}
}
\caption{Cost and Latency Comparison for Filter Queries.}
\vspace{-2.5em}
\label{table: filter}
\end{table*}

\subsection{Evaluation of Logical Optimization (RQ4)}

\begin{figure}
    \centering
    \includegraphics[width=0.9\linewidth]{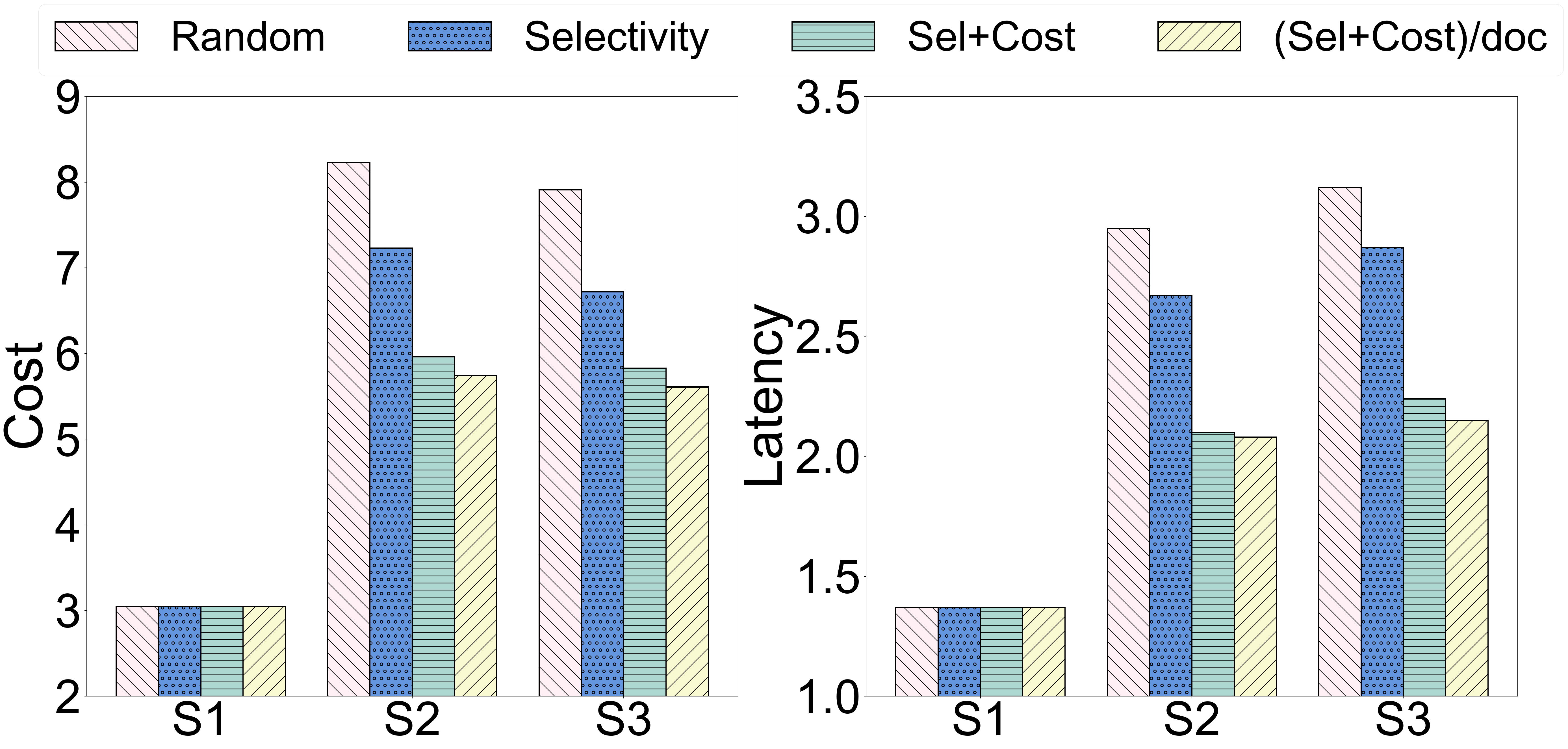}
    \vspace{-1.3em}
    \caption{Evaluation of Filter Reordering.}
    \vspace{-1.5em}
    \label{fig:filter recodering}
\end{figure}

\noindent \textbf{Filter Reordering.} We compare with four filter reordering strategies as follows.
(1) \texttt{Random}: the filters are executed in random order; (2) \texttt{Selectivity}: the filters are ordered based on the selectivity; (3) \texttt{Sel+Cost} (\zendb): the filters are ordered based on both the selectivity and the estimated average cost of extracting each attribute from the sampled documents;  (4) \texttt{(Sel+Cost)/doc} (\quest): each document has its own plan considering the selectivity and the estimated extraction cost w.r.t. this document.
We compare the above strategies based on \quest. To evaluate sufficiently,  we vary the number of filters: S1 with one filter, S2 with 2-3 filters, and S3 with 4 or more. We execute five queries in each of these three categories on \nba.
In Figure~\ref{fig:filter recodering}, for queries in C1, the cost of all baselines is almost identical because there is only one filter and hence one order per query. For queries with more filters, these methods are ranked as follows by the LLMs cost: \texttt{(Sel+Cost)/doc} < \texttt{Sel+Cost} < \texttt{Selectivity} < \texttt{Random}. The first two strategies save more cost because they optimize the order considering the cost. \texttt{(Sel+Cost)/doc} is the most cost-effective because it provides fine-grained optimization for each individual document.

\begin{figure}
    \centering
    \includegraphics[width=0.8\linewidth]{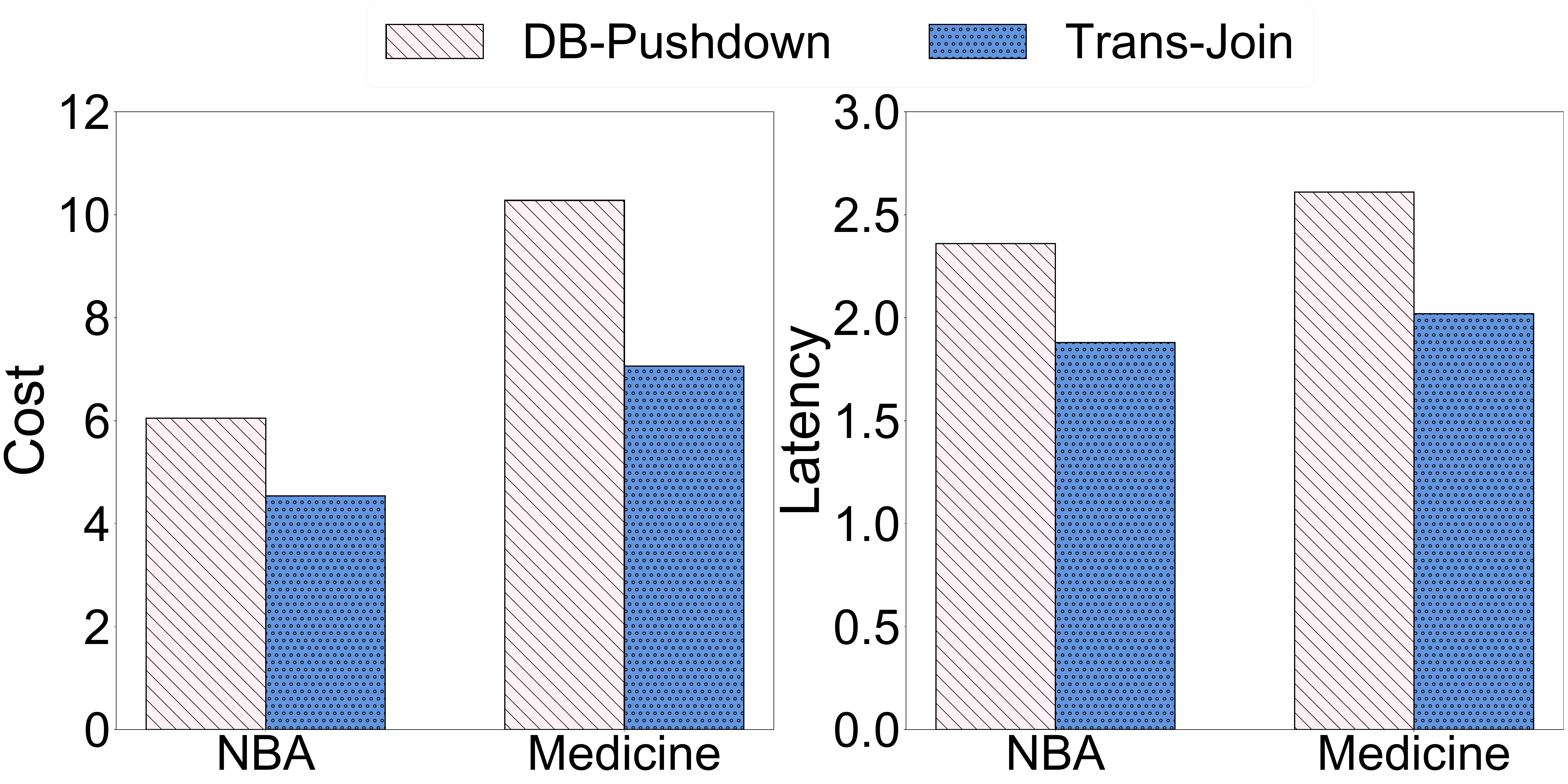}
    \vspace{-1.3em}
    \caption{Evaluation of Filter Pushdown.}
    \vspace{-1.6em}
    \label{fig:filter pushdown}
\end{figure}

\noindent \textbf{Filter Pushdown.} We compare with two strategies as follows. (1) \texttt{DB-Pushdown}: like in traditional databases, it pushes down filters respectively to the relevant tables before join. 
(2) \texttt{Trans-Join} (\quest): it transforms a join to a filter operation and orders the filters using the above (Sel+Cost)/doc strategy to reduce the cost. 
We compare the above strategies using 5 join queries with filters on \nba. We observe from Figure~\ref{fig:filter pushdown} that \texttt{Trans-Join} is more cost-effective than \texttt{DB-Pushdown} because given two tables to join, \texttt{Trans-Join} builds a cost model to judiciously determine which table will be extracted first and transformed to a filter on the other table, which provides the opportunity to run a join first if it incurs a smaller data extraction cost. 
%

\begin{figure}
    \centering
    \includegraphics[width=0.8\linewidth]{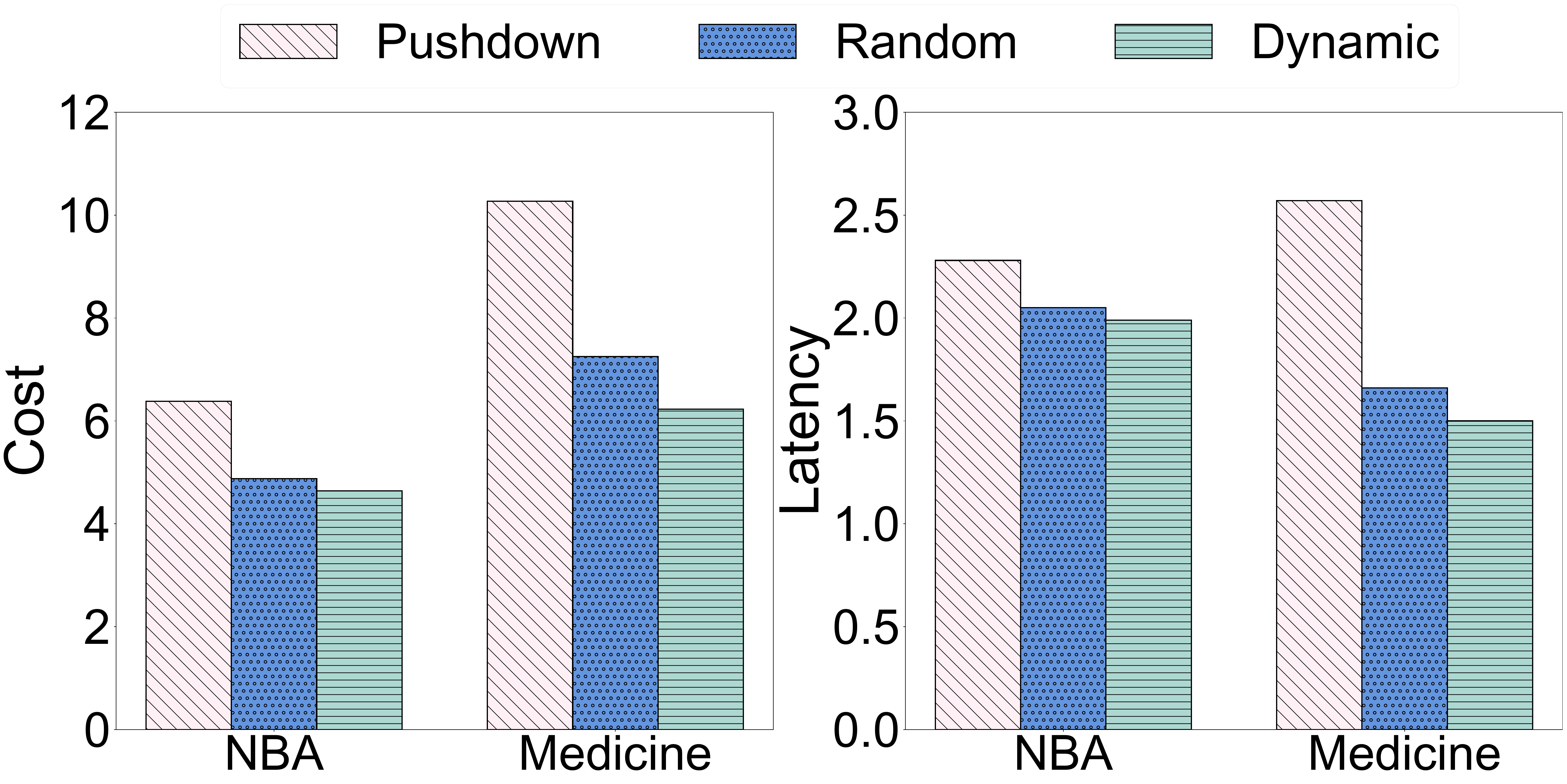}
    \vspace{-1.3em}
    \caption{Evaluation of Join Order.}
    \vspace{-2.1em}
    \label{fig:join order}
\end{figure}
\noindent \textbf{Join order.} We evaluate three strategies as follows. (1) \texttt{Pushdown}: all filters are pushed down first, and then join is performed; (2) \texttt{Random}:  tables (document subsets) are joined in random order, with each pair of tables joined using \texttt{Trans-Join}; (3) \texttt{Dynamic} (\zendb, \quest): it uses a cost model to dynamically identify two tables to join in a left-deep manner, with each pair of tables joined using \texttt{Trans-Join}. We compare the above strategies using five join queries involving more than three tables on \nba. We observe from Figure~\ref{fig:join order} that \texttt{Dynamic} saves much cost because it dynamically selects the join operation that leads to the lowest cost. 

\subsection{Evaluation of Physical Optimization (RQ5)}

\noindent \textbf{Model Selection.}
\begin{figure}
    \centering
    \includegraphics[width=0.9\linewidth]{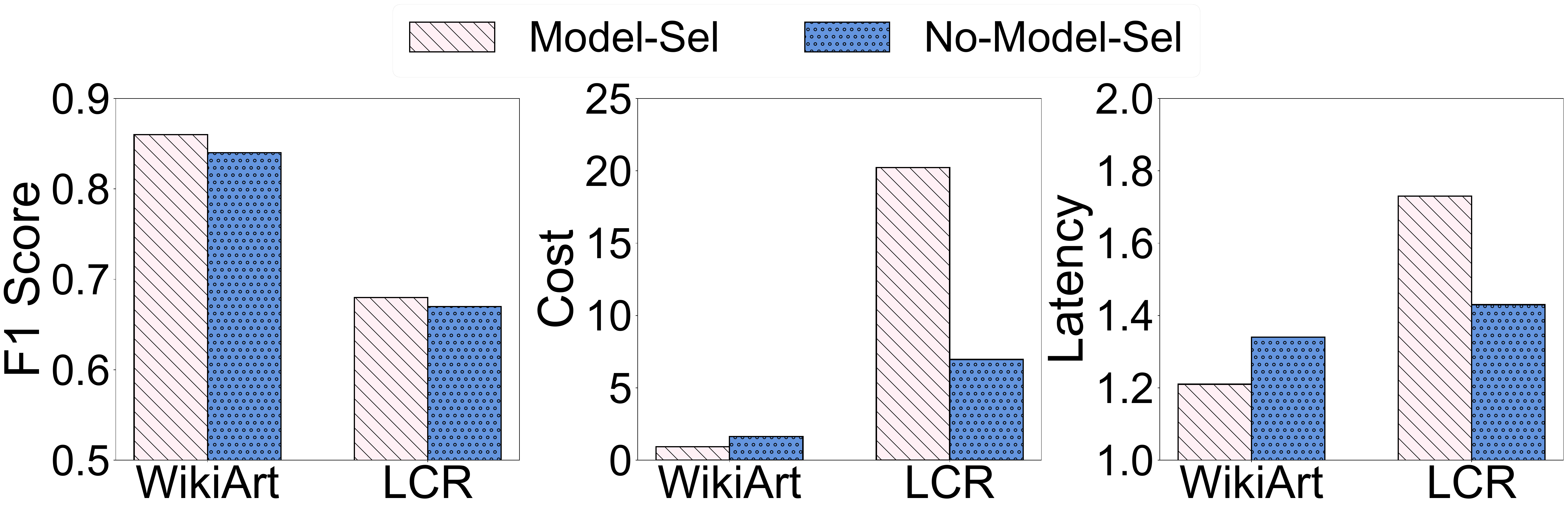}
    \vspace{-1.3em}
    \caption{Evaluation of Model Selection.}
    \vspace{-1.5em}
    \label{fig:model sel}
\end{figure}
We evaluate two strategies in \pz as follows.
(1) \texttt{Model-Sel}: We define a set of candidate models (\texttt{GPT-4.1-nano}, \texttt{GPT-4.1-mini}, and \texttt{GPT-4.1}) and set the selection objective as ``minimize cost while achieving the target accuracy.''
(2) \texttt{no-Model-Sel}: The system always uses a specific model (\texttt{GPT-4.1-mini}) for all queries.
The user has to specify a desired accuracy. In our experiments, we set this value as the average accuracy obtained by running five queries with \texttt{GPT-4.1-mini} on each dataset (\art and \legal).
We evaluate both strategies using the same set of queries on each dataset, including both the simple (\art) and challenging (\legal) datasets. The cost here is calculated by multiplying the number of tokens by the model’s price per token.
As shown in Figure~\ref{fig:model sel}, on \art, \texttt{Model-Sel} exceeds the target accuracy, while also reducing cost by using a cheaper model. This is achieved by assigning \texttt{GPT-4.1} to harder attributes and \texttt{GPT-4.1-nano} to easier ones, while \art has more easy attribute than hard attributes.
On the challenging \legal dataset, \texttt{Model-Sel} outperforms \texttt{No-Model-Sel} in accuracy, but with higher cost due to more frequently using \texttt{GPT-4.1}.

\noindent \textbf{Model Cascades.}
\begin{figure}
    \centering
    \includegraphics[width=0.9\linewidth]{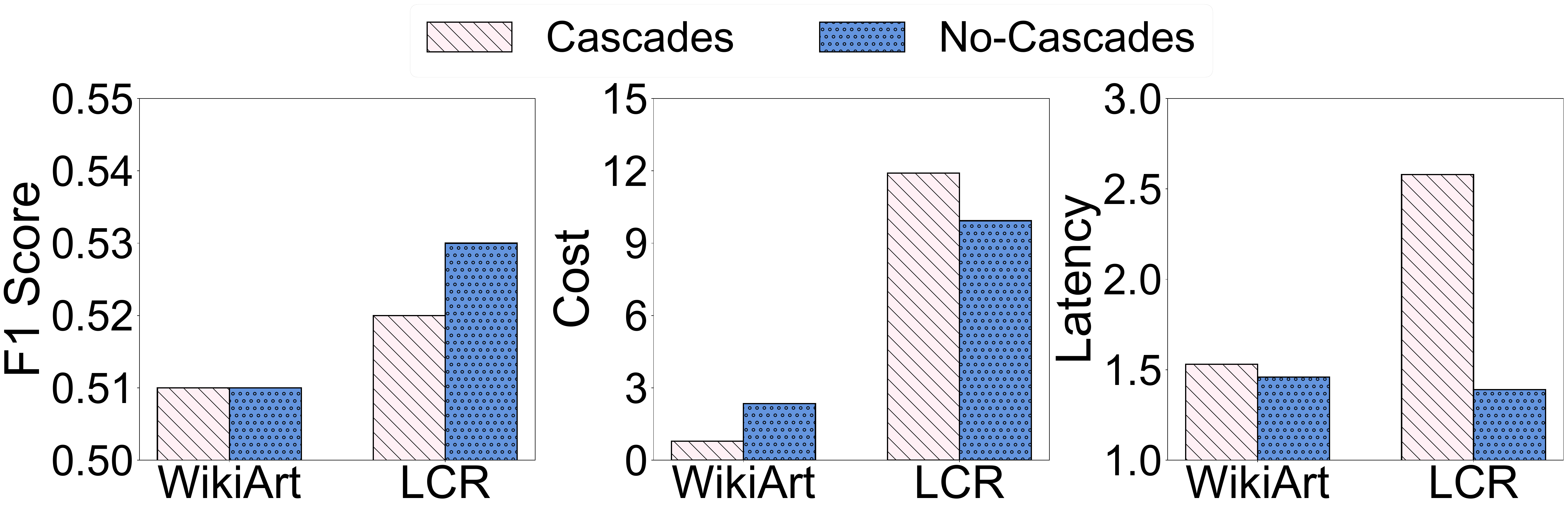}
    \vspace{-1.3em}
    \caption{Evaluation of Model Cascades.}
    \vspace{-1.4em}
    \label{fig:model cascade}
\end{figure}
We evaluate two strategies in \lotus as follows.
(1) \texttt{Cascades}: we construct a model cascade using two models, \texttt{GPT-4.1-nano} and \texttt{GPT-4.1-mini}. For each query, the system first uses the lower-cost \texttt{GPT-4.1-nano} to execute the queries. If the accuracy does not meet the desired accuracy, the query is then forwarded to the larger \texttt{GPT-4.1-mini} for further processing. 
(2) \texttt{No-Cascades}: The system uses only a single model, \texttt{GPT-4.1-mini} to process all queries.
We evaluate both strategies on the same set of queries for each dataset and use the same cost and target accuracy settings as the above model selection experiment.
As shown in Figure~\ref{fig:model cascade}, on \art, \texttt{Cascades} achieves the target accuracy while reducing cost by primarily using the less expensive \texttt{GPT-4.1-nano}. This is because, for most attributes in \art, the cheaper model is sufficient.
On the challenging \legal dataset, \texttt{Cascades} perform similar as \texttt{No-Cascades}, but incurs higher costs. This is because almost every attribute eventually requires \texttt{GPT-4.1-mini}, and all documents have to be first processed by \texttt{GPT-4.1-nano} before being fed to \texttt{GPT-4.1-mini}.

\noindent \textbf{Parallel Execution.}
\begin{figure}
    \centering
    \includegraphics[width=0.8\linewidth]{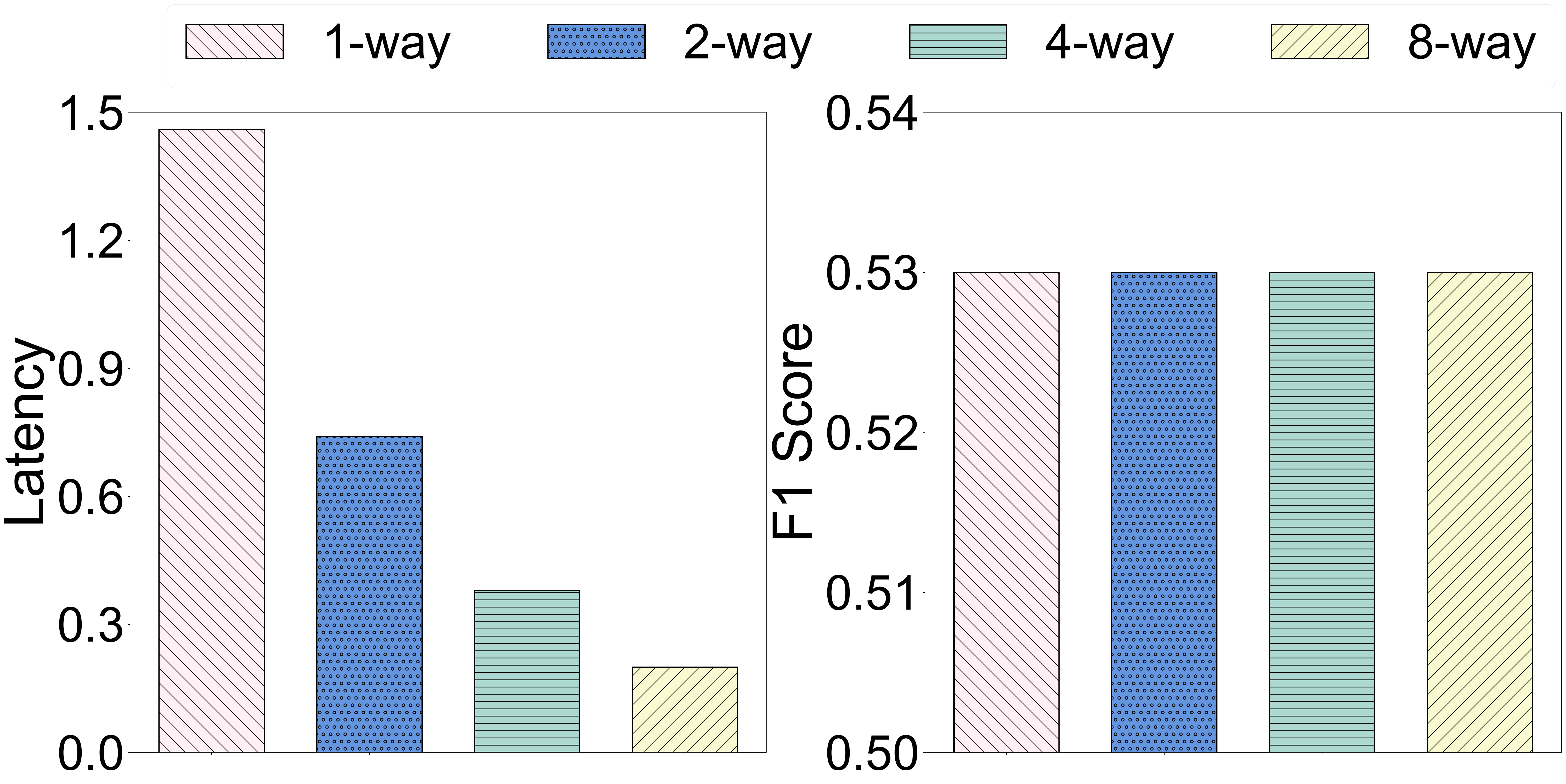}
    \vspace{-1.3em}
    \caption{Evaluation of Parallel Execution.}
    \vspace{-1.5em}
    \label{fig:parallel execution}
\end{figure}
We evaluate two strategies on \med using five queries as follows. 
(1) \texttt{Parallel}: employing parallel execution with different levels of parallelism, specifically with 2-way, 4-way, and 8-way thread parallelism.
(2) \texttt{no-Parallel}: using a sequential approach without any parallelism.
As shown in Figure ~\ref{fig:parallel execution}, we observe that increasing the levels of parallelism leads to a proportional reduction in execution time. For example, with 2-way, 4-way, and 8-way thread parallelism, the execution time is reduced from 1.46 seconds (\texttt{no-Parallel}) to 0.74, 0.38, and 0.2 seconds, respectively, while the F1 score remains unchanged at 0.53.

\begin{center}
\vspace{-1em}
\fcolorbox{black}{gray!10}{\parbox{\linewidth}{Opportunity II: 
Currently, there is no system supporting end-to-end query optimization, including all above logical and physical optimization strategies, which would be a promising direction to achieve high-efficacy query execution.
}}
\end{center}

\begin{figure}
    \centering
    \includegraphics[width=0.8\linewidth]{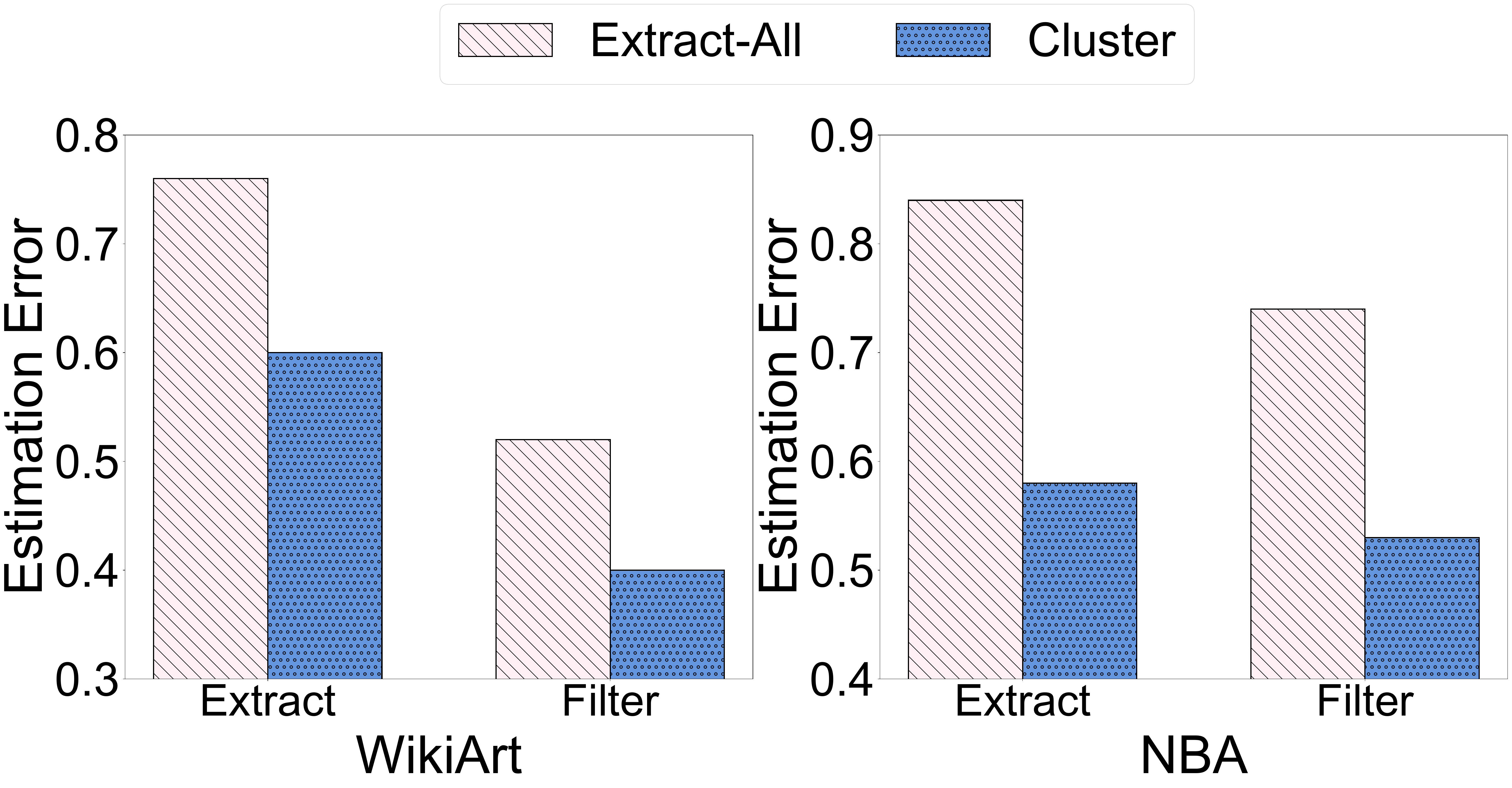}
    \vspace{-1.3em}
    \caption{Evaluation of  Aggregation.}
    \vspace{-1.6em}
    \label{fig:agg optimization}
\end{figure}
\noindent \textbf{Optimization for Aggregation.}
We use the estimation error as the metric following~\cite{uqe}, defined as the absolute difference between the estimated and true values divided by the true value. We evaluate two strategies on \nba using five queries as follows. 
(1) \texttt{Extract-All}: it extracts the attributes involved in Groupby and Aggregation and then executes the query.
(2) \texttt{Cluster}: it clusters the attribute-related chunks based on the attribute in Groupby, extracts the attribute in Aggregation within each cluster and executes the query.
We observe from Figure~\ref{fig:agg optimization} that \texttt{Cluster} is more cost-effective than \texttt{Extract-All} because it does not need to extract the Groupby attribute, i.e., assuming the documents in each cluster share the same attribute value. However, the relative error is very high because it is hard to have a high-quality clustering simply based on the chunk embeddings.

\begin{figure}
    \centering
    \includegraphics[width=0.9\linewidth]{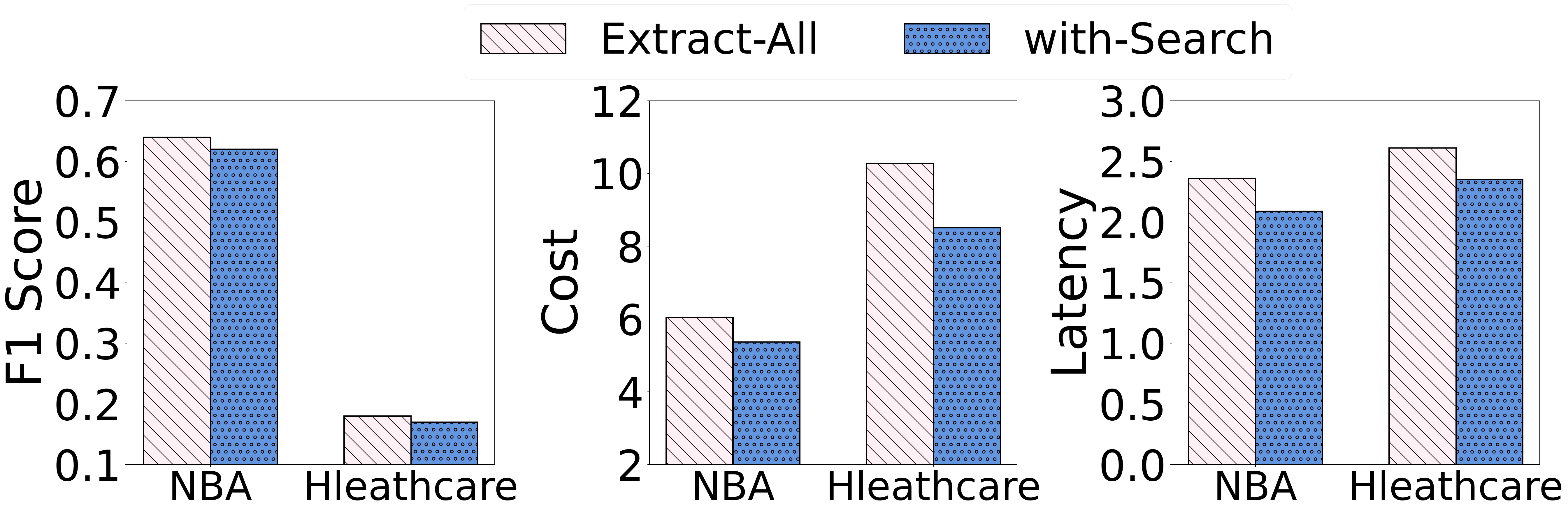}
    \vspace{-1.5em}
    \caption{Evaluation of Join Optimization.}
    \vspace{-1.6em}
    \label{fig:join optimization}
\end{figure}
\noindent \textbf{Optimization for Join.}
We compare with two strategies on \nba using five queries as follows. 
(1) \texttt{Extract-All}:  it extracts the join key attribute from the two tables and join them.
(2) \texttt{with-Search}: it extracts each value of the join key attribute from one table, leverages it as a semantic search key to prune many documents in the other table and then join.
We compare the above strategies using five join queries. We can observe from Figure~\ref{fig:join optimization} that \texttt{with-Search} is more cost-effective than \texttt{Extract-All} because many documents in the other table are pruned without extraction. However, the accuracy is  low because it  incorrectly prunes documents that can be joined.
%

\begin{center}
\vspace{-1em}
\fcolorbox{black}{gray!10}{\parbox{\linewidth}{Opportunity III: 
Another promising direction is to develop more effective strategies to align the embeddings of attributes with the embeddings of their chunks.  In this way, the clustering in aggregation and pruning in join can be more accurate, and thereby the costs can be reduced more.
}}
\end{center}

\subsection{Ablation Studies}
\label{subsec:ablation study}

\begin{figure}
    \centering
    \includegraphics[width=0.9\linewidth]{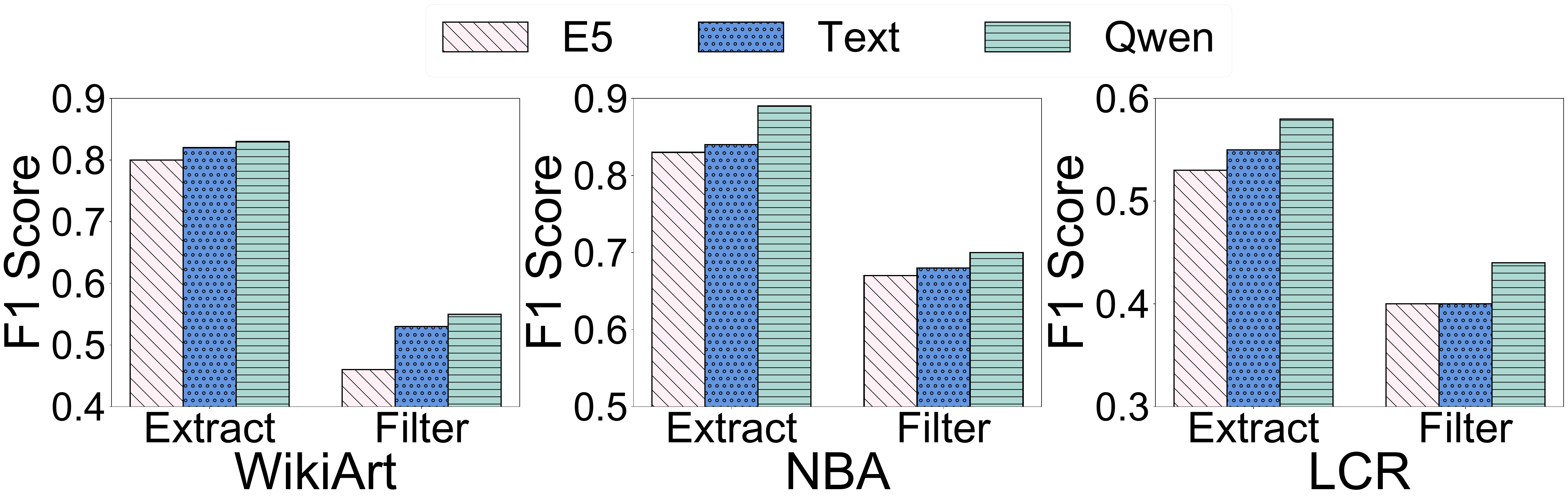}
    \vspace{-1.5em}
    \caption{Ablation Study of Embedding Models.}
    \vspace{-2.2em}
    \label{fig:embedding}
\end{figure}
\noindent \textbf{Ablation Study of Embedding.}
We evaluate different embedding models to analyze whether using a better embedding model can improve the system performance. 
We utilize \texttt{
multilingual-e5-larg\\e}\texttt{(E5)}~\cite{E5}, \texttt{text-embedding-3}(\texttt{Text})~\cite{text-embedding-3} and \texttt{Qwen3-Embdding}\texttt{(Q\\wen)}~\cite{qwen3embedding}. 
On  MTEB benchmark~\cite{mteb}, their performance is ranked as follows: \texttt{Qwen} $>$ \texttt{Text} $>$ \texttt{E5}.
We evaluate the performance of different embedding models by replacing the embedding model in \quest and testing on the \art, \nba, and \legal datasets, each with five queries each 
As shown in Figure~\ref{fig:embedding}, system performance improves as the quality of the embedding model increases. This is because better embedding models provide more accurate semantic representations, resulting in more precise chunk retrieval, and thus improve overall performance.

\begin{figure}
    \centering
    \includegraphics[width=0.9\linewidth]{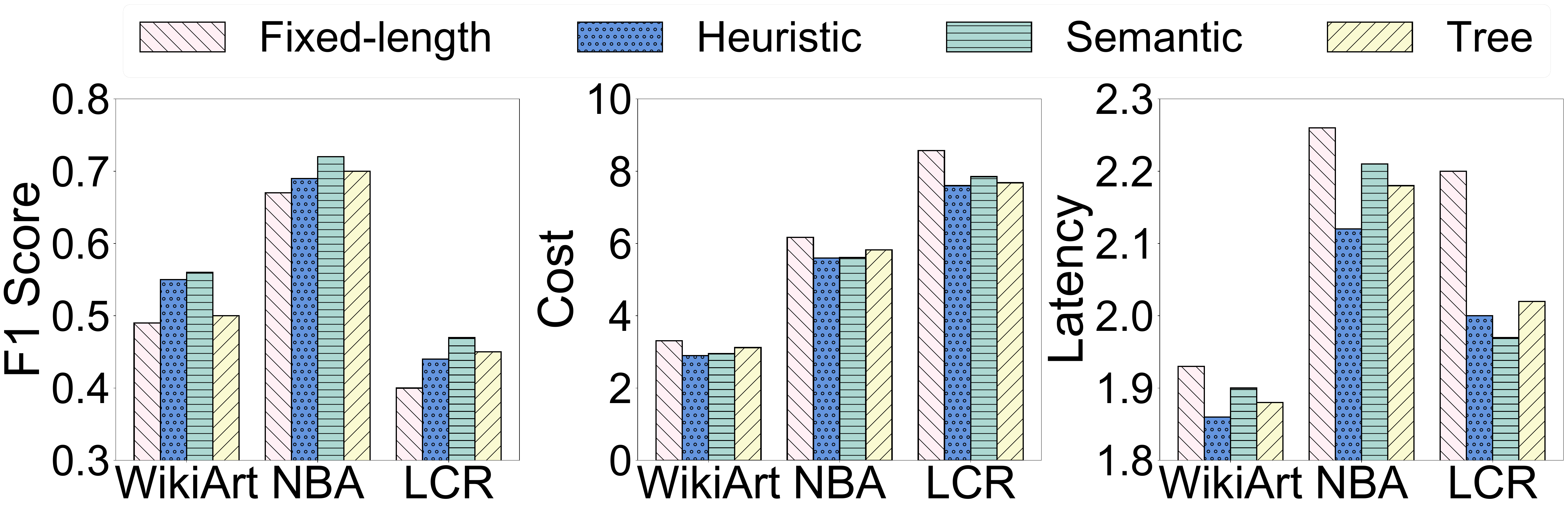}
    \vspace{-0.5em}
    \caption{Ablation Study of Chunking Strategies.}
    \vspace{-0.5em}
    \label{fig:chunk}
\end{figure}
\noindent \textbf{Ablation Study of Chunking Strategies.} 
We evaluate how various chunking strategies affect system performance. 
We use the following chunking strategies:
(1) \texttt{Fixed-length}~\cite{fixed-chunk}: chunks are set to a fixed size of 512 tokens.
(2) \texttt{Heuristic}~\cite{grammer-chunk}: grammar-based chunking with a maximum chunk size of 512 tokens.
(3) \texttt{Semantic} (\quest): semantic-based chunking with a maximum length of 512 tokens and a minimum length of 128 tokens.
(4) \texttt{Tree} (\zendb):  an SHT tree to split the documents.
Details for each chunking strategy are in Appendix. We test performance by replacing the chunking strategy in \quest and conducting five queries on datasets \art, \nba, and \legal.
As shown in Figure~\ref{fig:chunk}, \texttt{Semantic} always performs the best since it captures the rich semantic representations and thus leads to a better embedding similarity than other strategies.
\texttt{Tree} outperforms \texttt{Heuristic} on \nba and \legal because it preserves more tokens per chunk, providing richer in-context information, although this comes at a higher cost and latency.
\texttt{Fixed-length} performs the worst, as it may split complete sentences into different chunks, resulting in incomplete information.

\noindent \textbf{Ablation Study of Different LLMs.}
\begin{figure}
    \centering
    \includegraphics[width=0.7\linewidth]{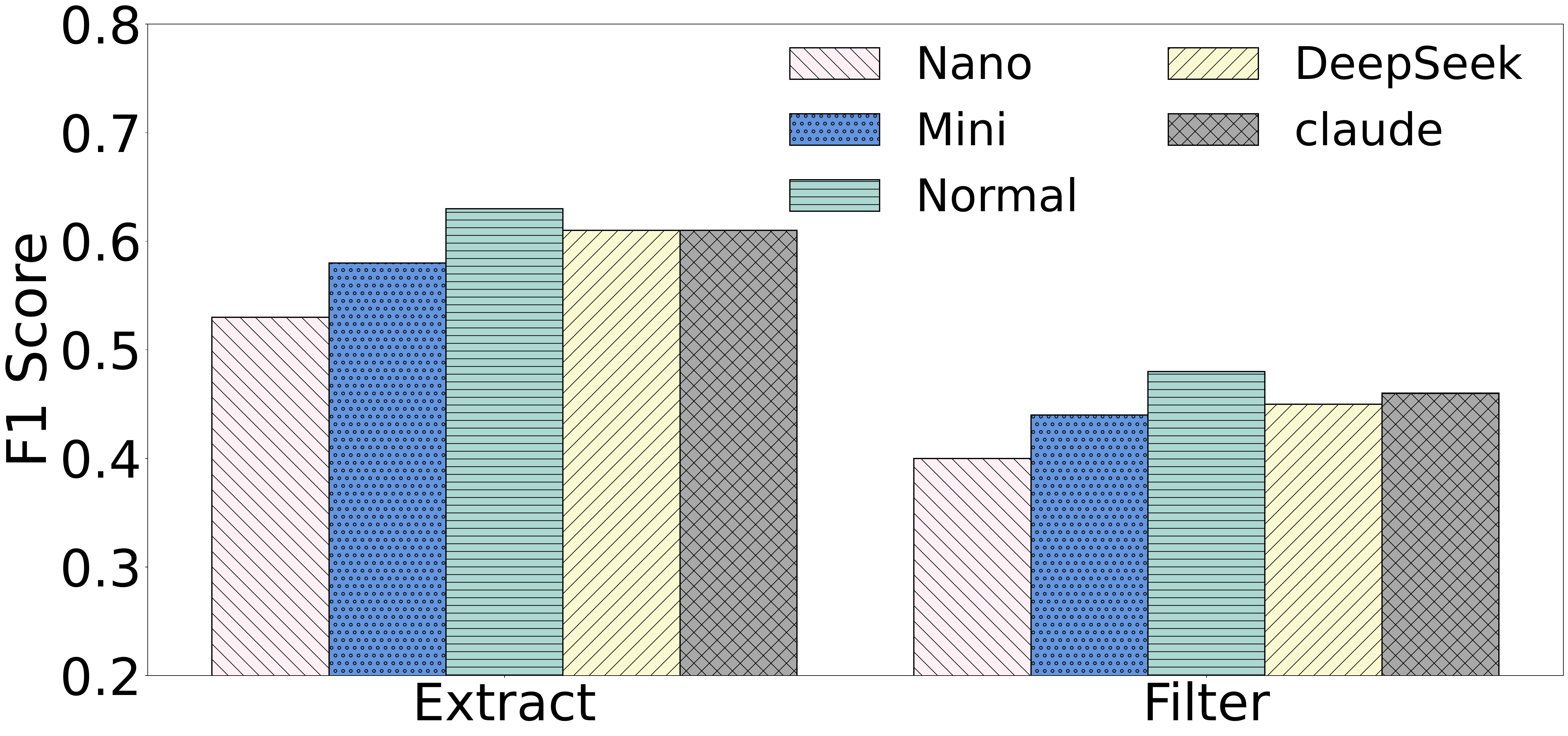}
    \vspace{-1em}
    \caption{Ablation Study of Different LLMs.}
    \vspace{-2em}
    \label{fig:model}
\end{figure}
We evaluate the impact of  different types of  LLMs (\texttt{GPT-4.1-nano (Nano)}, \texttt{GPT-4.1-mini (Mini)}, \texttt{GPT-4.1} (Normal), \texttt{Deepseek-V3 (Deepseek)}, and \texttt{Claude-\\sonnet-4 (Claude)}) on system performance to assess whether stronger LLMs lead to improved results.
We evaluate the performance of various LLMs by replacing the LLM API in \quest and testing each model on five queries from the \legal dataset. Figure~\ref{fig:model} presents the results, with performance ranked as: \texttt{GPT-4.1} $>$ \texttt{Claude-sonnet-4} $>$ \texttt{Deepseek-V3} $>$ \texttt{GPT-4.1-mini} $>$ \texttt{GPT-4.1-nano}, which closely aligns with the expected capability of these LLMs~\cite{vellum-leaderboard}.

\section{related work}\label{sec: related work}
In Section~\ref{sec: system}, we have reviewed the key components of existing LLM-powered UDA systems. This section focuses on the pro and cons of these systems and analyzes their overall performance. In addition, we review techniques for data extraction, which is a key operation in such systems.

\noindent \textbf{LLM-powered Unstructured Data Analysis Systems.}
%
%
Lotus~\cite{lotus} introduces semantic operators for unstructured data processing, including indexing, extraction, filtering, joining capabilities that enable the construction of complex analytical pipelines. It provides optimized physical implementation for each operator but lacks logical optimizations. Its experimental evaluation is limited to small datasets with few queries --five queries corresponding to five tasks.
%
Palimpzest~\cite{pz} offers libraries for users to write declarative Python code to analyze unstructured data. It optimizes the logical plan of each program via filter reordering based on selectivities. It optimizes the physical plan mainly by selecting LLMs for a task based on user preference. However, the dataset used in the evaluation is relatively small --1,149 documents in total.  
DocETL\cite{docetl} focuses on improving the accuracy of UDA using a multi-agent strategy, but it does not conduct logical optimization to save cost. 
Its evaluation is performed on five datasets, with only one representative query per dataset.
ZenDB\cite{zendb} uses semantic hierarchical trees for identifying relevant document sections and applies filter ordering and predicate pushdown for optimization. However, it requires well-structured documents and is evaluated on just 221 documents and 27 queries, with no publicly available ground truth.
UQE~\cite{uqe} provides SQL-like analysis with sampling-based aggregation, while
CAESURA~\cite{CAESURA} decomposes queries into operators handling data in different modalities. Both systems, however, are evaluated on limited-size datasets. For example, UQE provides one query and the dataset contains only 1,000 shot emails.
Similarly, early systems~\cite{symphony,thalamusdb,thorne2021from,urban2023towards} employ LLMs for document analysis but overlook cost optimization, which is a critical concern given the computational expense of LLM inference, and lack a thorough benchmarking.

\noindent \textbf{LLM-powered Data Extraction.}
Extracting information rom unstructured sources has evolved from rule-based methods~\cite{lee2013attribute,niklaus2018survey,saha2018open,saha2017bootstrapping} to modern deep learning approaches. OpenIE6~\cite{OpenIE6} employs iterative grid labeling for triple extraction, while MacroIE~\cite{MacroIE} uses BERT-based encoders to identify entity relationships. However, these methods struggle with implicit relationships and complex document structures. 
LLM-based systems seek to resolve these issues. \eva~\cite{eva} generates extraction code through LLMs, balancing cost and quality through weak supervision. 

Recent works focus on Pre-trained Language Models (PLMs) for data extraction, including DebertaV3~\cite{debertav3}, which uses QA task pre-training for text information extraction, Text-to-Table\cite{Text-to-Table} using a sequence-to-sequence model for text-to-table conversion, STable\cite{stable} with a permutation-based decoder for flexibility, and ODIE~\cite{ODIE} applying LoRA to fine-tune a LLaMA-7B model for unstructured text extraction. These techniques can enhance the cost-effectiveness and quality of UDA systems.

\section{conclusion}\label{sec: conclusion}

In this paper, we build a comprehensive benchmark for unstructured data analysis powered by LLMs. We collect five unstructured datasets with diverse characteristics, based on which we define a number of significant attributes and accurately label their values with human efforts. We also construct hundreds of meaningful queries with various analytical operators. Finally, we implement existing systems, run queries over the labeled datasets to test their performance and conduct in-depth analysis.

\bibliographystyle{ACM-Reference-Format}
\bibliography{references}


\begin{thebibliography}{38}


\ifx \showCODEN    \undefined \def \showCODEN     #1{\unskip}     \fi
\ifx \showDOI      \undefined \def \showDOI       #1{#1}\fi
\ifx \showISBNx    \undefined \def \showISBNx     #1{\unskip}     \fi
\ifx \showISBNxiii \undefined \def \showISBNxiii  #1{\unskip}     \fi
\ifx \showISSN     \undefined \def \showISSN      #1{\unskip}     \fi
\ifx \showLCCN     \undefined \def \showLCCN      #1{\unskip}     \fi
\ifx \shownote     \undefined \def \shownote      #1{#1}          \fi
\ifx \showarticletitle \undefined \def \showarticletitle #1{#1}   \fi
\ifx \showURL      \undefined \def \showURL       {\relax}        \fi
\providecommand\bibfield[2]{#2}
\providecommand\bibinfo[2]{#2}
\providecommand\natexlab[1]{#1}
\providecommand\showeprint[2][]{arXiv:#2}

\bibitem[\protect\citeauthoryear{??}{vel}{2024}]%
        {vellum-leaderboard}
 \bibinfo{year}{2024}\natexlab{}.
\newblock \bibinfo{title}{Vellum AI LLM Leaderboard}.
\newblock \bibinfo{howpublished}{\url{https://www.vellum.ai/llm-leaderboard}}.
\newblock
\newblock
\shownote{\url{https://www.vellum.ai/llm-leaderboard}.}


\bibitem[\protect\citeauthoryear{Arora, Yang, Eyuboglu, Narayan, Hojel,
  Trummer, and Ré}{Arora et~al\mbox{.}}{2025}]%
        {eva}
\bibfield{author}{\bibinfo{person}{Simran Arora}, \bibinfo{person}{Brandon
  Yang}, \bibinfo{person}{Sabri Eyuboglu}, \bibinfo{person}{Avanika Narayan},
  \bibinfo{person}{Andrew Hojel}, \bibinfo{person}{Immanuel Trummer}, {and}
  \bibinfo{person}{Christopher Ré}.} \bibinfo{year}{2025}\natexlab{}.
\newblock \showarticletitle{Language Models Enable Simple Systems for
  Generating Structured Views of Heterogeneous Data Lakes}.
\newblock
\showeprint[arxiv]{2304.09433}~[cs.CL]
\urldef\tempurl%
\url{https://arxiv.org/abs/2304.09433}
\showURL{%
\tempurl}


\bibitem[\protect\citeauthoryear{{Art‑.org}}{{Art‑.org}}{2025}]%
        {artorg2025}
\bibfield{author}{\bibinfo{person}{{Art‑.org}}.}
  \bibinfo{year}{2025}\natexlab{}.
\newblock \bibinfo{booktitle}{\emph{{Art‑.org – Artists and Artworks
  (19th–21st C.)}}}.
\newblock
\urldef\tempurl%
\url{https://www.art-.org/}
\showURL{%
\tempurl}


\bibitem[\protect\citeauthoryear{{Australasian Legal Information Institute
  (AustLII)}}{{Australasian Legal Information Institute (AustLII)}}{[n.d.]}]%
        {austlii}
\bibfield{author}{\bibinfo{person}{{Australasian Legal Information Institute
  (AustLII)}}.} \bibinfo{year}{[n.d.]}\natexlab{}.
\newblock \bibinfo{title}{{AustLII – Australasian Legal Information
  Institute}}.
\newblock \bibinfo{howpublished}{\url{https://www.austlii.edu.au/}}.
\newblock


\bibitem[\protect\citeauthoryear{Bhat, Rudat, Spiekermann, and
  Flores-Herr}{Bhat et~al\mbox{.}}{2025}]%
        {fixed-chunk}
\bibfield{author}{\bibinfo{person}{Sinchana~Ramakanth Bhat},
  \bibinfo{person}{Max Rudat}, \bibinfo{person}{Jannis Spiekermann}, {and}
  \bibinfo{person}{Nicolas Flores-Herr}.} \bibinfo{year}{2025}\natexlab{}.
\newblock \bibinfo{title}{Rethinking Chunk Size For Long-Document Retrieval: A
  Multi-Dataset Analysis}.
\newblock
\newblock
\showeprint[arxiv]{2505.21700}~[cs.IR]
\urldef\tempurl%
\url{https://arxiv.org/abs/2505.21700}
\showURL{%
\tempurl}


\bibitem[\protect\citeauthoryear{Chen, Gu, Cao, Fan, Madden, and Tang}{Chen
  et~al\mbox{.}}{2023}]%
        {symphony}
\bibfield{author}{\bibinfo{person}{Zui Chen}, \bibinfo{person}{Zihui Gu},
  \bibinfo{person}{Lei Cao}, \bibinfo{person}{Ju Fan}, \bibinfo{person}{Sam
  Madden}, {and} \bibinfo{person}{Nan Tang}.} \bibinfo{year}{2023}\natexlab{}.
\newblock \bibinfo{title}{Symphony: Towards Natural Language Query Answering
  over Multi-modal Data Lakes}.
\newblock
\newblock
\urldef\tempurl%
\url{https://www.cidrdb.org/cidr2023/papers/p51-chen.pdf}
\showURL{%
\tempurl}


\bibitem[\protect\citeauthoryear{Dai, Wang, Wan, Dai, Yang, Nova, Yin,
  Phothilimthana, Sutton, and Schuurmans}{Dai et~al\mbox{.}}{2024}]%
        {uqe}
\bibfield{author}{\bibinfo{person}{Hanjun Dai}, \bibinfo{person}{Bethany~Yixin
  Wang}, \bibinfo{person}{Xingchen Wan}, \bibinfo{person}{Bo Dai},
  \bibinfo{person}{Sherry Yang}, \bibinfo{person}{Azade Nova},
  \bibinfo{person}{Pengcheng Yin}, \bibinfo{person}{Phitchaya~Mangpo
  Phothilimthana}, \bibinfo{person}{Charles Sutton}, {and}
  \bibinfo{person}{Dale Schuurmans}.} \bibinfo{year}{2024}\natexlab{}.
\newblock \bibinfo{title}{UQE: A Query Engine for Unstructured Databases}.
\newblock
\newblock
\showeprint[arxiv]{2407.09522}~[cs.DB]
\urldef\tempurl%
\url{https://arxiv.org/abs/2407.09522}
\showURL{%
\tempurl}


\bibitem[\protect\citeauthoryear{{Enterprise RAG Challenge}}{{Enterprise RAG
  Challenge}}{[n.d.]}]%
        {enterpriserag2024}
\bibfield{author}{\bibinfo{person}{{Enterprise RAG Challenge}}.}
  \bibinfo{year}{[n.d.]}\natexlab{}.
\newblock \bibinfo{title}{{Enterprise RAG Challenge}}.
\newblock \bibinfo{howpublished}{\url{https://rag.abdullin.com/}}.
\newblock
\newblock
\shownote{Accessed 17 July 2025.}


\bibitem[\protect\citeauthoryear{He, Gao, and Chen}{He et~al\mbox{.}}{2023}]%
        {debertav3}
\bibfield{author}{\bibinfo{person}{Pengcheng He}, \bibinfo{person}{Jianfeng
  Gao}, {and} \bibinfo{person}{Weizhu Chen}.} \bibinfo{year}{2023}\natexlab{}.
\newblock \bibinfo{title}{DeBERTaV3: Improving DeBERTa using ELECTRA-Style
  Pre-Training with Gradient-Disentangled Embedding Sharing}.
\newblock
\newblock
\showeprint[arxiv]{2111.09543}~[cs.CL]
\urldef\tempurl%
\url{https://arxiv.org/abs/2111.09543}
\showURL{%
\tempurl}


\bibitem[\protect\citeauthoryear{Jiao, Zhong, Li, Zhao, Ouyang, Ji, and
  Han}{Jiao et~al\mbox{.}}{2023}]%
        {ODIE}
\bibfield{author}{\bibinfo{person}{Yizhu Jiao}, \bibinfo{person}{Ming Zhong},
  \bibinfo{person}{Sha Li}, \bibinfo{person}{Ruining Zhao},
  \bibinfo{person}{Siru Ouyang}, \bibinfo{person}{Heng Ji}, {and}
  \bibinfo{person}{Jiawei Han}.} \bibinfo{year}{2023}\natexlab{}.
\newblock \bibinfo{title}{Instruct and Extract: Instruction Tuning for
  On-Demand Information Extraction}.
\newblock
\newblock
\showeprint[arxiv]{2310.16040}~[cs.CL]
\urldef\tempurl%
\url{https://arxiv.org/abs/2310.16040}
\showURL{%
\tempurl}


\bibitem[\protect\citeauthoryear{Jo and Trummer}{Jo and Trummer}{2024}]%
        {thalamusdb}
\bibfield{author}{\bibinfo{person}{Saehan Jo} {and} \bibinfo{person}{Immanuel
  Trummer}.} \bibinfo{year}{2024}\natexlab{}.
\newblock \bibinfo{title}{ThalamusDB: Approximate Query Processing on
  Multi-Modal Data}.
\newblock , \bibinfo{numpages}{26}~pages.
\newblock
\urldef\tempurl%
\url{https://dl.acm.org/doi/10.1145/3654989}
\showURL{%
\tempurl}


\bibitem[\protect\citeauthoryear{Kanimozhi and Venkatesan}{Kanimozhi and
  Venkatesan}{2015}]%
        {idc}
\bibfield{author}{\bibinfo{person}{K.~V. Kanimozhi} {and} \bibinfo{person}{M.
  Venkatesan}.} \bibinfo{year}{2015}\natexlab{}.
\newblock \showarticletitle{Unstructured Data Analysis--A Survey}.
\newblock \bibinfo{journal}{\emph{International Journal of Advanced Research in
  Computer and Communication Engineering}} \bibinfo{volume}{4},
  \bibinfo{number}{3} (\bibinfo{year}{2015}), \bibinfo{pages}{223--225}.
\newblock


\bibitem[\protect\citeauthoryear{Kolluru, Adlakha, Aggarwal, Mausam, and
  Chakrabarti}{Kolluru et~al\mbox{.}}{2020}]%
        {OpenIE6}
\bibfield{author}{\bibinfo{person}{Keshav Kolluru}, \bibinfo{person}{Vaibhav
  Adlakha}, \bibinfo{person}{Samarth Aggarwal}, \bibinfo{person}{Mausam}, {and}
  \bibinfo{person}{Soumen Chakrabarti}.} \bibinfo{year}{2020}\natexlab{}.
\newblock \showarticletitle{OpenIE6: Iterative Grid Labeling and Coordination
  Analysis for Open Information Extraction}. In
  \bibinfo{booktitle}{\emph{Proceedings of the 2020 Conference on Empirical
  Methods in Natural Language Processing (EMNLP)}}.
  \bibinfo{publisher}{Association for Computational Linguistics},
  \bibinfo{address}{Online}, \bibinfo{pages}{3748--3761}.
\newblock
\urldef\tempurl%
\url{https://doi.org/10.18653/v1/2020.emnlp-main.306}
\showDOI{\tempurl}


\bibitem[\protect\citeauthoryear{Kwon, Li, Zhuang, Sheng, Zheng, Yu, Gonzalez,
  Zhang, and Stoica}{Kwon et~al\mbox{.}}{[n.d.]}]%
        {vllm}
\bibfield{author}{\bibinfo{person}{Woosuk Kwon}, \bibinfo{person}{Zhuohan Li},
  \bibinfo{person}{Siyuan Zhuang}, \bibinfo{person}{Ying Sheng},
  \bibinfo{person}{Lianmin Zheng}, \bibinfo{person}{CodyHao Yu},
  \bibinfo{person}{JosephE Gonzalez}, \bibinfo{person}{Hao Zhang}, {and}
  \bibinfo{person}{Ion Stoica}.} \bibinfo{year}{[n.d.]}\natexlab{}.
\newblock \showarticletitle{Efficient Memory Management for Large Language
  Model Serving with PagedAttention}.
\newblock  (\bibinfo{year}{[n.\,d.]}).
\newblock


\bibitem[\protect\citeauthoryear{Lee, Wang, Wang, and Hwang}{Lee
  et~al\mbox{.}}{2013}]%
        {lee2013attribute}
\bibfield{author}{\bibinfo{person}{Taesung Lee}, \bibinfo{person}{Zhongyuan
  Wang}, \bibinfo{person}{Haixun Wang}, {and} \bibinfo{person}{Seung-won
  Hwang}.} \bibinfo{year}{2013}\natexlab{}.
\newblock \showarticletitle{Attribute extraction and scoring: A probabilistic
  approach}. In \bibinfo{booktitle}{\emph{2013 IEEE 29th International
  Conference on Data Engineering (ICDE)}}. \bibinfo{pages}{194--205}.
\newblock
\urldef\tempurl%
\url{https://doi.org/10.1109/ICDE.2013.6544825}
\showDOI{\tempurl}


\bibitem[\protect\citeauthoryear{Lin, Hulsebos, Ma, Shankar, Zeigham,
  Parameswaran, and Wu}{Lin et~al\mbox{.}}{2024}]%
        {zendb}
\bibfield{author}{\bibinfo{person}{Yiming Lin}, \bibinfo{person}{Madelon
  Hulsebos}, \bibinfo{person}{Ruiying Ma}, \bibinfo{person}{Shreya Shankar},
  \bibinfo{person}{Sepanta Zeigham}, \bibinfo{person}{Aditya~G. Parameswaran},
  {and} \bibinfo{person}{Eugene Wu}.} \bibinfo{year}{2024}\natexlab{}.
\newblock \bibinfo{title}{Towards Accurate and Efficient Document Analytics
  with Large Language Models}.
\newblock
\newblock
\showeprint[arxiv]{2405.04674}~[cs.DB]
\urldef\tempurl%
\url{https://arxiv.org/abs/2405.04674}
\showURL{%
\tempurl}


\bibitem[\protect\citeauthoryear{Liu, Russo, Cafarella, Cao, Chen, Chen,
  Franklin, Kraska, Madden, and Vitagliano}{Liu et~al\mbox{.}}{2024}]%
        {pz}
\bibfield{author}{\bibinfo{person}{Chunwei Liu}, \bibinfo{person}{Matthew
  Russo}, \bibinfo{person}{Michael Cafarella}, \bibinfo{person}{Lei Cao},
  \bibinfo{person}{Peter~Baille Chen}, \bibinfo{person}{Zui Chen},
  \bibinfo{person}{Michael Franklin}, \bibinfo{person}{Tim Kraska},
  \bibinfo{person}{Samuel Madden}, {and} \bibinfo{person}{Gerardo Vitagliano}.}
  \bibinfo{year}{2024}\natexlab{}.
\newblock \bibinfo{title}{A Declarative System for Optimizing AI Workloads}.
\newblock
\newblock
\showeprint[arxiv]{2405.14696}~[cs.CL]
\urldef\tempurl%
\url{https://arxiv.org/abs/2405.14696}
\showURL{%
\tempurl}


\bibitem[\protect\citeauthoryear{Loper and Bird}{Loper and Bird}{2002}]%
        {NLTK}
\bibfield{author}{\bibinfo{person}{Edward Loper} {and} \bibinfo{person}{Steven
  Bird}.} \bibinfo{year}{2002}\natexlab{}.
\newblock \bibinfo{title}{NLTK: The Natural Language Toolkit}.
\newblock
\newblock
\showeprint[arxiv]{cs/0205028}~[cs.CL]
\urldef\tempurl%
\url{https://arxiv.org/abs/cs/0205028}
\showURL{%
\tempurl}


\bibitem[\protect\citeauthoryear{Muennighoff, Tazi, Magne, and
  Reimers}{Muennighoff et~al\mbox{.}}{2023}]%
        {mteb}
\bibfield{author}{\bibinfo{person}{Niklas Muennighoff},
  \bibinfo{person}{Nouamane Tazi}, \bibinfo{person}{Loïc Magne}, {and}
  \bibinfo{person}{Nils Reimers}.} \bibinfo{year}{2023}\natexlab{}.
\newblock \bibinfo{title}{MTEB: Massive Text Embedding Benchmark}.
\newblock
\newblock
\showeprint[arxiv]{2210.07316}~[cs.CL]
\urldef\tempurl%
\url{https://arxiv.org/abs/2210.07316}
\showURL{%
\tempurl}


\bibitem[\protect\citeauthoryear{{National Basketball Association –
  Wikipedia}}{{National Basketball Association – Wikipedia}}{[n.d.]}]%
        {nba_wikipedia}
\bibfield{author}{\bibinfo{person}{{National Basketball Association –
  Wikipedia}}.} \bibinfo{year}{[n.d.]}\natexlab{}.
\newblock \bibinfo{title}{{NBA – Wikipedia}}.
\newblock
  \bibinfo{howpublished}{\url{https://en.wikipedia.org/wiki/National_Basketball_Association}}.
\newblock
\newblock
\shownote{Accessed 17 July 2025.}


\bibitem[\protect\citeauthoryear{Neelakantan, Xu, Puri, Radford, Han, Tworek,
  Yuan, Tezak, Kim, Hallacy, Heidecke, Shyam, Power, Nekoul, Sastry, Krueger,
  Schnurr, Such, Hsu, Thompson, Khan, Sherbakov, Jang, Welinder, and
  Weng}{Neelakantan et~al\mbox{.}}{2022}]%
        {text-embedding-3}
\bibfield{author}{\bibinfo{person}{Arvind Neelakantan}, \bibinfo{person}{Tao
  Xu}, \bibinfo{person}{Raul Puri}, \bibinfo{person}{Alec Radford},
  \bibinfo{person}{Jesse~Michael Han}, \bibinfo{person}{Jerry Tworek},
  \bibinfo{person}{Qiming Yuan}, \bibinfo{person}{Nikolas Tezak},
  \bibinfo{person}{Jong~Wook Kim}, \bibinfo{person}{Chris Hallacy},
  \bibinfo{person}{Johannes Heidecke}, \bibinfo{person}{Pranav Shyam},
  \bibinfo{person}{Boris Power}, \bibinfo{person}{Tyna~Eloundou Nekoul},
  \bibinfo{person}{Girish Sastry}, \bibinfo{person}{Gretchen Krueger},
  \bibinfo{person}{David Schnurr}, \bibinfo{person}{Felipe~Petroski Such},
  \bibinfo{person}{Kenny Hsu}, \bibinfo{person}{Madeleine Thompson},
  \bibinfo{person}{Tabarak Khan}, \bibinfo{person}{Toki Sherbakov},
  \bibinfo{person}{Joanne Jang}, \bibinfo{person}{Peter Welinder}, {and}
  \bibinfo{person}{Lilian Weng}.} \bibinfo{year}{2022}\natexlab{}.
\newblock \bibinfo{title}{Text and Code Embeddings by Contrastive
  Pre-Training}.
\newblock
\newblock
\showeprint[arxiv]{2201.10005}~[cs.CL]
\urldef\tempurl%
\url{https://arxiv.org/abs/2201.10005}
\showURL{%
\tempurl}


\bibitem[\protect\citeauthoryear{Niklaus, Cetto, Freitas, and
  Handschuh}{Niklaus et~al\mbox{.}}{2018}]%
        {niklaus2018survey}
\bibfield{author}{\bibinfo{person}{Christina Niklaus},
  \bibinfo{person}{Matthias Cetto}, \bibinfo{person}{André Freitas}, {and}
  \bibinfo{person}{Siegfried Handschuh}.} \bibinfo{year}{2018}\natexlab{}.
\newblock \bibinfo{title}{A Survey on Open Information Extraction}.
\newblock
\newblock
\showeprint[arxiv]{1806.05599}~[cs.CL]
\urldef\tempurl%
\url{https://arxiv.org/abs/1806.05599}
\showURL{%
\tempurl}


\bibitem[\protect\citeauthoryear{Patel, Jha, Pan, Gupta, Asawa, Guestrin, and
  Zaharia}{Patel et~al\mbox{.}}{2025}]%
        {lotus}
\bibfield{author}{\bibinfo{person}{Liana Patel}, \bibinfo{person}{Siddharth
  Jha}, \bibinfo{person}{Melissa Pan}, \bibinfo{person}{Harshit Gupta},
  \bibinfo{person}{Parth Asawa}, \bibinfo{person}{Carlos Guestrin}, {and}
  \bibinfo{person}{Matei Zaharia}.} \bibinfo{year}{2025}\natexlab{}.
\newblock \bibinfo{title}{Semantic Operators: A Declarative Model for Rich,
  AI-based Data Processing}.
\newblock
\newblock
\showeprint[arxiv]{2407.11418}~[cs.DB]
\urldef\tempurl%
\url{https://arxiv.org/abs/2407.11418}
\showURL{%
\tempurl}


\bibitem[\protect\citeauthoryear{Pietruszka, Turski, Łukasz Borchmann, Dwojak,
  Nowakowska, Szyndler, Jurkiewicz, and Łukasz Garncarek}{Pietruszka
  et~al\mbox{.}}{2024}]%
        {stable}
\bibfield{author}{\bibinfo{person}{Michał Pietruszka},
  \bibinfo{person}{Michał Turski}, \bibinfo{person}{Łukasz Borchmann},
  \bibinfo{person}{Tomasz Dwojak}, \bibinfo{person}{Gabriela Nowakowska},
  \bibinfo{person}{Karolina Szyndler}, \bibinfo{person}{Dawid Jurkiewicz},
  {and} \bibinfo{person}{Łukasz Garncarek}.} \bibinfo{year}{2024}\natexlab{}.
\newblock \bibinfo{title}{STable: Table Generation Framework for
  Encoder-Decoder Models}.
\newblock
\newblock
\showeprint[arxiv]{2206.04045}~[cs.CL]
\urldef\tempurl%
\url{https://arxiv.org/abs/2206.04045}
\showURL{%
\tempurl}


\bibitem[\protect\citeauthoryear{Qiu, Wu, Zhang, Lin, Wang, Zhang, Wang, and
  Xie}{Qiu et~al\mbox{.}}{2024}]%
        {mmbench}
\bibfield{author}{\bibinfo{person}{Pengcheng Qiu}, \bibinfo{person}{Chaoyi Wu},
  \bibinfo{person}{Xiaoman Zhang}, \bibinfo{person}{Weixiong Lin},
  \bibinfo{person}{Haicheng Wang}, \bibinfo{person}{Ya Zhang},
  \bibinfo{person}{Yanfeng Wang}, {and} \bibinfo{person}{Weidi Xie}.}
  \bibinfo{year}{2024}\natexlab{}.
\newblock \bibinfo{title}{Towards Building Multilingual Language Model for
  Medicine}.
\newblock
\newblock
\showeprint[arxiv]{2402.13963}~[cs.CL]
\urldef\tempurl%
\url{https://arxiv.org/abs/2402.13963}
\showURL{%
\tempurl}


\bibitem[\protect\citeauthoryear{Saha and Mausam}{Saha and Mausam}{2018}]%
        {saha2018open}
\bibfield{author}{\bibinfo{person}{Swarnadeep Saha} {and}
  \bibinfo{person}{Mausam}.} \bibinfo{year}{2018}\natexlab{}.
\newblock \bibinfo{title}{Open Information Extraction from Conjunctive
  Sentences}.
\newblock , \bibinfo{numpages}{2288--2299}~pages.
\newblock
\urldef\tempurl%
\url{https://aclanthology.org/C18-1194/}
\showURL{%
\tempurl}


\bibitem[\protect\citeauthoryear{Saha, Pal, and Mausam}{Saha
  et~al\mbox{.}}{2017}]%
        {saha2017bootstrapping}
\bibfield{author}{\bibinfo{person}{Swarnadeep Saha}, \bibinfo{person}{Harinder
  Pal}, {and} \bibinfo{person}{Mausam}.} \bibinfo{year}{2017}\natexlab{}.
\newblock \bibinfo{title}{Bootstrapping for Numerical Open IE}.
\newblock , \bibinfo{numpages}{317--323}~pages.
\newblock
\urldef\tempurl%
\url{https://doi.org/10.18653/v1/P17-2050}
\showURL{%
\tempurl}


\bibitem[\protect\citeauthoryear{Schwaber‑Cohen and Patel}{Schwaber‑Cohen
  and Patel}{2025}]%
        {grammer-chunk}
\bibfield{author}{\bibinfo{person}{Roie Schwaber‑Cohen} {and}
  \bibinfo{person}{Arjun Patel}.} \bibinfo{year}{2025}\natexlab{}.
\newblock \bibinfo{title}{Chunking Strategies for LLM Applications}.
\newblock \bibinfo{howpublished}{Pinecone Blog}.
\newblock
\urldef\tempurl%
\url{https://www.pinecone.io/learn/chunking-strategies-for-llm-applications/}
\showURL{%
\tempurl}
\newblock
\shownote{Retrieved 17 July 2025 from Pinecone website.}


\bibitem[\protect\citeauthoryear{Shankar, Chambers, Shah, Parameswaran, and
  Wu}{Shankar et~al\mbox{.}}{2025}]%
        {docetl}
\bibfield{author}{\bibinfo{person}{Shreya Shankar}, \bibinfo{person}{Tristan
  Chambers}, \bibinfo{person}{Tarak Shah}, \bibinfo{person}{Aditya~G.
  Parameswaran}, {and} \bibinfo{person}{Eugene Wu}.}
  \bibinfo{year}{2025}\natexlab{}.
\newblock \bibinfo{title}{DocETL: Agentic Query Rewriting and Evaluation for
  Complex Document Processing}.
\newblock
\newblock
\showeprint[arxiv]{2410.12189}~[cs.DB]
\urldef\tempurl%
\url{https://arxiv.org/abs/2410.12189}
\showURL{%
\tempurl}


\bibitem[\protect\citeauthoryear{Sun, Deng, Chai, Jin, Guo, Han, Yuan, Wang,
  and Cao}{Sun et~al\mbox{.}}{2025}]%
        {quest}
\bibfield{author}{\bibinfo{person}{Zhaoze Sun}, \bibinfo{person}{Qiyan Deng},
  \bibinfo{person}{Chengliang Chai}, \bibinfo{person}{Kaisen Jin},
  \bibinfo{person}{Xinyu Guo}, \bibinfo{person}{Han Han}, \bibinfo{person}{Ye
  Yuan}, \bibinfo{person}{Guoren Wang}, {and} \bibinfo{person}{Lei Cao}.}
  \bibinfo{year}{2025}\natexlab{}.
\newblock \bibinfo{title}{QUEST: Query Optimization in Unstructured Document
  Analysis}.
\newblock
\newblock
\showeprint[arxiv]{2507.06515}~[cs.DB]
\urldef\tempurl%
\url{https://arxiv.org/abs/2507.06515}
\showURL{%
\tempurl}


\bibitem[\protect\citeauthoryear{Thorne, Yazdani, Saeidi, Silvestri, Riedel,
  and Halevy}{Thorne et~al\mbox{.}}{2021}]%
        {thorne2021from}
\bibfield{author}{\bibinfo{person}{James Thorne}, \bibinfo{person}{Majid
  Yazdani}, \bibinfo{person}{Marzieh Saeidi}, \bibinfo{person}{Fabrizio
  Silvestri}, \bibinfo{person}{Sebastian Riedel}, {and} \bibinfo{person}{Alon
  Halevy}.} \bibinfo{year}{2021}\natexlab{}.
\newblock \bibinfo{title}{From Natural Language Processing to Neural
  Databases}.
\newblock , \bibinfo{numpages}{1033--1039}~pages.
\newblock
\urldef\tempurl%
\url{https://doi.org/10.14778/3447689.3447706}
\showURL{%
\tempurl}


\bibitem[\protect\citeauthoryear{Urban and Binnig}{Urban and Binnig}{2023}]%
        {urban2023towards}
\bibfield{author}{\bibinfo{person}{Matthias Urban} {and}
  \bibinfo{person}{Carsten Binnig}.} \bibinfo{year}{2023}\natexlab{}.
\newblock \bibinfo{title}{Towards Multi-Modal DBMSs for Seamless Querying of
  Texts and Tables}.
\newblock
\newblock
\showeprint[arxiv]{2304.13559}~[cs.DB]
\urldef\tempurl%
\url{https://arxiv.org/abs/2304.13559}
\showURL{%
\tempurl}


\bibitem[\protect\citeauthoryear{Urban and Binnig}{Urban and Binnig}{2024}]%
        {CAESURA}
\bibfield{author}{\bibinfo{person}{Matthias Urban} {and}
  \bibinfo{person}{Carsten Binnig}.} \bibinfo{year}{2024}\natexlab{}.
\newblock \showarticletitle{CAESURA: Language Models as Multi-Modal Query
  Planners}. In \bibinfo{booktitle}{\emph{14th Conference on Innovative Data
  Systems Research, {CIDR} 2024, Chaminade, CA, USA, January 14-17, 2024}}.
  \bibinfo{publisher}{www.cidrdb.org}.
\newblock
\urldef\tempurl%
\url{https://www.cidrdb.org/cidr2024/papers/p14-urban.pdf}
\showURL{%
\tempurl}


\bibitem[\protect\citeauthoryear{Wang, Xu, Zhao, Ouyang, Wu, Zhao, Xu, Liu, Qu,
  Shang, Zhang, Wei, Sui, Li, Shi, Qiao, Lin, and He}{Wang
  et~al\mbox{.}}{2024a}]%
        {mineru}
\bibfield{author}{\bibinfo{person}{Bin Wang}, \bibinfo{person}{Chao Xu},
  \bibinfo{person}{Xiaomeng Zhao}, \bibinfo{person}{Linke Ouyang},
  \bibinfo{person}{Fan Wu}, \bibinfo{person}{Zhiyuan Zhao},
  \bibinfo{person}{Rui Xu}, \bibinfo{person}{Kaiwen Liu}, \bibinfo{person}{Yuan
  Qu}, \bibinfo{person}{Fukai Shang}, \bibinfo{person}{Bo Zhang},
  \bibinfo{person}{Liqun Wei}, \bibinfo{person}{Zhihao Sui},
  \bibinfo{person}{Wei Li}, \bibinfo{person}{Botian Shi}, \bibinfo{person}{Yu
  Qiao}, \bibinfo{person}{Dahua Lin}, {and} \bibinfo{person}{Conghui He}.}
  \bibinfo{year}{2024}\natexlab{a}.
\newblock \bibinfo{title}{MinerU: An Open-Source Solution for Precise Document
  Content Extraction}.
\newblock
\newblock
\showeprint[arxiv]{2409.18839}~[cs.CV]
\urldef\tempurl%
\url{https://arxiv.org/abs/2409.18839}
\showURL{%
\tempurl}


\bibitem[\protect\citeauthoryear{Wang, Yang, Huang, Yang, Majumder, and
  Wei}{Wang et~al\mbox{.}}{2024b}]%
        {E5}
\bibfield{author}{\bibinfo{person}{Liang Wang}, \bibinfo{person}{Nan Yang},
  \bibinfo{person}{Xiaolong Huang}, \bibinfo{person}{Linjun Yang},
  \bibinfo{person}{Rangan Majumder}, {and} \bibinfo{person}{Furu Wei}.}
  \bibinfo{year}{2024}\natexlab{b}.
\newblock \bibinfo{title}{Multilingual E5 Text Embeddings: A Technical Report}.
\newblock
\newblock
\showeprint[arxiv]{2402.05672}~[cs.CL]
\urldef\tempurl%
\url{https://arxiv.org/abs/2402.05672}
\showURL{%
\tempurl}


\bibitem[\protect\citeauthoryear{Wu, Zhang, and Li}{Wu et~al\mbox{.}}{2022}]%
        {Text-to-Table}
\bibfield{author}{\bibinfo{person}{Xueqing Wu}, \bibinfo{person}{Jiacheng
  Zhang}, {and} \bibinfo{person}{Hang Li}.} \bibinfo{year}{2022}\natexlab{}.
\newblock \bibinfo{title}{Text-to-Table: A New Way of Information Extraction}.
\newblock
\newblock
\showeprint[arxiv]{2109.02707}~[cs.CL]
\urldef\tempurl%
\url{https://arxiv.org/abs/2109.02707}
\showURL{%
\tempurl}


\bibitem[\protect\citeauthoryear{Yu, Wang, Liu, Zhu, Sun, and Wang}{Yu
  et~al\mbox{.}}{2021}]%
        {MacroIE}
\bibfield{author}{\bibinfo{person}{Bowen Yu}, \bibinfo{person}{Yucheng Wang},
  \bibinfo{person}{Tingwen Liu}, \bibinfo{person}{Hongsong Zhu},
  \bibinfo{person}{Limin Sun}, {and} \bibinfo{person}{Bin Wang}.}
  \bibinfo{year}{2021}\natexlab{}.
\newblock \showarticletitle{Maximal Clique Based Non-Autoregressive Open
  Information Extraction}. In \bibinfo{booktitle}{\emph{Proceedings of the 2021
  Conference on Empirical Methods in Natural Language Processing}},
  \bibfield{editor}{\bibinfo{person}{Marie-Francine Moens},
  \bibinfo{person}{Xuanjing Huang}, \bibinfo{person}{Lucia Specia}, {and}
  \bibinfo{person}{Scott Wen-tau Yih}} (Eds.). \bibinfo{publisher}{Association
  for Computational Linguistics}, \bibinfo{address}{Online and Punta Cana,
  Dominican Republic}, \bibinfo{pages}{9696--9706}.
\newblock
\urldef\tempurl%
\url{https://doi.org/10.18653/v1/2021.emnlp-main.764}
\showDOI{\tempurl}


\bibitem[\protect\citeauthoryear{Zhang, Li, Long, Zhang, Lin, Yang, Xie, Yang,
  Liu, Lin, Huang, and Zhou}{Zhang et~al\mbox{.}}{2025}]%
        {qwen3embedding}
\bibfield{author}{\bibinfo{person}{Yanzhao Zhang}, \bibinfo{person}{Mingxin
  Li}, \bibinfo{person}{Dingkun Long}, \bibinfo{person}{Xin Zhang},
  \bibinfo{person}{Huan Lin}, \bibinfo{person}{Baosong Yang},
  \bibinfo{person}{Pengjun Xie}, \bibinfo{person}{An Yang},
  \bibinfo{person}{Dayiheng Liu}, \bibinfo{person}{Junyang Lin},
  \bibinfo{person}{Fei Huang}, {and} \bibinfo{person}{Jingren Zhou}.}
  \bibinfo{year}{2025}\natexlab{}.
\newblock \showarticletitle{Qwen3 Embedding: Advancing Text Embedding and
  Reranking Through Foundation Models}.
\newblock \bibinfo{journal}{\emph{arXiv preprint arXiv:2506.05176}}
  (\bibinfo{year}{2025}).
\newblock


\end{thebibliography}


\end{document}